\PassOptionsToPackage{table,xcdraw}{xcolor}
\documentclass[acmsmall]{acmart}
\def\BibTeX{{\rm B\kern-.05em{\sc i\kern-.025em b}\kern-.08emT\kern-.1667em\lower.7ex\hbox{E}\kern-.125emX}}

\setcopyright{acmcopyright}
\acmJournal{TOSEM}
\acmYear{2021}\acmVolume{01}\acmNumber{1}\acmArticle{000}\acmMonth{1}
\acmDOI{10.1145/1122445.1122456}

\DeclareMathOperator*{\argmax}{argmax}
\usepackage{xcolor}
\usepackage{graphicx}
\usepackage{algorithm}
\usepackage{textcomp}
\usepackage{listings}
\usepackage{comment}
\usepackage{multirow}
\usepackage{tabularx}
\usepackage{tcolorbox}
\usepackage{listings}
\usepackage{array}
\usepackage{balance}
\usepackage{seqsplit}

\newboolean{showcomments}
\setboolean{showcomments}{false}
\ifthenelse{\boolean{showcomments}}
  {\newcommand{\nb}[2]{
  \fbox{\bfseries\sffamily\scriptsize#1}
     {\sf\small$\blacktriangleright$\textit{\textcolor{red}{#2}}$\blacktriangleleft$}
   }
  }
  {\newcommand{\nb}[2]{}
   
  }

\definecolor{codegreen}{rgb}{0,0.6,0}
\definecolor{codegray}{rgb}{0.5,0.5,0.5}
\definecolor{codepurple}{rgb}{0.58,0,0.82}
\definecolor{backcolour}{rgb}{0.95,0.95,0.92}

\lstdefinestyle{mystyle}{
    backgroundcolor=\color{backcolour},   
    commentstyle=\color{codegreen},
    keywordstyle=\color{purple},
    numberstyle=\tiny\color{codegray},
    stringstyle=\color{magenta},
    basicstyle=\ttfamily\footnotesize,
    breakatwhitespace=false,         
    breaklines=true,                 
    captionpos=b,                    
    keepspaces=true,                 
    numbers=left,                    
    numbersep=5pt,                  
    showspaces=false,                
    showstringspaces=false,
    showtabs=false,
    tabsize=2
}

\newcommand\toolname{ARES}

\begin{document}

\title{Deep Reinforcement Learning for Black-Box Testing of Android Apps}

\author{Andrea Romdhana}
\email{andrea.romdhana@dibris.unige.it}
\orcid{0000-0003-4651-8713}
\affiliation{%
  \institution{DIBRIS - Università degli Studi di Genova, FBK-ICT, Security \& Trust Unit}
}
\author{Alessio Merlo}
\email{alessio.merlo@unige.it}
\orcid{0000-0002-2272-2376}
\affiliation{%
  \institution{DIBRIS - Università degli Studi di Genova}
}
\author{Mariano Ceccato}
\email{mariano.ceccato@univr.it}
\orcid{0000-0001-7325-0316}
\affiliation{%
  \institution{Università di Verona}
}
\author{Paolo Tonella}
\email{paolo.tonella@usi.ch}
\orcid{0000-0003-3088-0339}
\affiliation{%
  \institution{Università della Svizzera italiana}
}

\begin{abstract}
The state space of Android apps is huge and its thorough exploration during testing remains a major challenge. In fact, the best exploration strategy is highly dependent on the features of the app under test. Reinforcement Learning (RL) is a machine learning technique that learns the optimal strategy to solve a task by trial and error, guided by positive or negative reward, rather than by explicit supervision. Deep RL is a recent extension of RL that takes advantage of the learning capabilities of neural networks. Such capabilities make Deep RL suitable for complex exploration spaces such as the one of Android apps. However, state of the art, publicly available tools only support basic, tabular RL.
We have developed \toolname{}, a Deep RL approach for black-box testing of Android apps. Experimental results show that it achieves higher coverage and fault revelation than the  baselines, which include state of the art tools, such as TimeMachine and Q-Testing. We also investigated qualitatively the reasons behind such performance and we have identified the  key features of Android apps that make Deep RL particularly effective on them to be the presence of chained and blocking activities.
\end{abstract}


\maketitle
\section{Introduction}
\label{sec:introduction}

The complexity of mobile applications (hereafter, apps) keeps growing, as apps provide always more advanced services to the users. Nonetheless, it is of utmost importance that they work properly once they are published on app markets as most of their success/failure depend on the user's evaluation. Therefore, an effective testing phase is fundamental to minimize the likelihood of app failures during execution.  
However, automated testing of mobile apps is still an open problem and the complexity of current apps makes their exploration trickier than in the past, as they can contain states that are difficult to reach and events that are hard to trigger. 

There exist several approaches to automated testing of mobile apps that aim at maximizing  code coverage and  bug detection during testing. Random testing strategies \cite{Monkey, machiry2013dynodroid} stimulate the App Under Test (AUT) by producing pseudo-random events. However, random exploration with no guidance may get stuck when dealing with complex transitions.
Model-Based strategies \cite{amalfitano2012using, DBLP:conf/sigsoft/SuMCWYYPLS17, gu2019practical} extract test cases from navigation models built by means of static or dynamic analysis. If the model accurately reflects the AUT, a deep exploration can be achieved. Nonetheless, automatically constructed models tend to be incomplete and inaccurate. 
Structural strategies \cite{anand2012automated, gao2018android, mahmood2014evodroid} generate coverage-oriented inputs using symbolic execution or evolutionary algorithms. These strategies are more powerful, since they are guided by a specific coverage target. However, they do not take advantage of past exploration successes to learn the most effective exploration strategy dynamically.

Reinforcement Learning (RL) is a machine learning approach that does not require a labeled training set as input, since the learning process is guided by the positive or negative reward experienced during the tentative execution of the task. Hence, it represents a way to dynamically build an optimal exploration strategy by taking advantage of the past successful or unsuccessful moves.

RL has been extensively applied to the problem of GUI and Android testing~\cite{mariani2012autoblacktest,qtest}. However, only the most basic form of RL (i.e., tabular RL) has been applied to testing problems so far. In tabular RL the value of the state-action associations is stored in a fixed table. The advent of Deep Neural Networks (DNN) replaced tabular approaches with deep learning ones, in which the action-value function is learned from the past positive and negative experiences made by one or more neural networks. When the state space to explore is extremely large (e.g. when an app has a big amount of widgets), Deep RL has proved to be substantially superior to tabular RL \cite{boyan1995generalization} \cite{riedmiller2005neural} \cite{li2017deep}. In this respect, we argue that the state space of Android apps is definitely a good candidate for a successful adoption of Deep RL instead of tabular RL for testing purposes.


This paper presents the first Deep RL approach, \toolname{}, for automated \textsl{black-box} testing of Android apps. \toolname{} uses a DNN to learn the best exploration strategy from previous attempts. Thanks to such DNN, it achieves high scalability, general applicability and the capability to handle complex app behaviors.

\toolname{} implements multiple Deep RL algorithms that come with a set of configurable, often critical, hyperparameters. 
To speed up both the selection of the most appropriate algorithm for the AUT and the fine tuning of its hyperparameters, we have developed another tool, FATE, which integrates with \toolname{}.

FATE is a simulation environment that supports fast assessment of Android testing algorithms by running \textit{synthetic} Android apps (i.e., abstract navigational models of real Android apps).
The execution of a testing session on a FATE synthetic app is on average 10 to 100 times faster than the  execution of the same  session on the corresponding real Android app. 

We applied \toolname{} to two benchmarks made by 41 and 68 Android apps, respectively. The first benchmark compares the performance of the \toolname{} algorithms, while the latter one evaluates \toolname{}  w.r.t. the state-of-the-art  testing tools for Android.

Experimental results confirmed the hypothesis that Deep RL outperforms tabular RL in the exploration of the state space of Android apps, as \toolname{} exposed the highest number of faults and obtained the highest code coverage. 
Furthermore, we carried out a qualitative analysis showing that the features of Android apps that make Deep RL particularly effective include, among others, the presence of concatenated activities and blocking activities, protected by authentication.

To sum up, this paper provides the following advancements to the state of the art:
\begin{itemize}
    \item \toolname{}, the first publicly available testing approach based on Deep Reinforcement Learning, released as open source;
    \item FATE, a simulation environment for fast experimentation of Android testing algorithms, also available as open source;
    \item A thorough empirical evaluation of the proposed approach, whose replication package is publicly available to the research community.
\end{itemize}

\section{Background}
\label{sec:background}

After a general overview on RL, this section presents in more detail Tabular RL  and Deep RL.

\subsection{Overview on Reinforcement Learning}

The objective of Reinforcement Learning~\cite{sutton} is to train an \textit{agent} that interacts with some environment to achieve a given {\em goal}. The agent is assumed to be capable of sensing the \textit{current state} of the \textit{environment}, and to receive a feedback signal, named \textit{reward}, each time the agent takes an \textit{action}. 

At each time step \textit{t}, the agent receives an observation $ x_t $, takes an action $ a_{t} $ that causes the transition of the environment from state $ s_t $ to state $ s_{t+1} $. The agent also receives a scalar reward $ R(x_t,a_t, x_{t+1}) $, that quantifies the goodness of the last transition.

For simplicity, let us assume $ x_t = s_t $ (in the general case, $x_t$ might be a subset of $s_t$). 
The behavior of an agent is represented by a \textit{policy} $ \pi $, i.e., a rule for making the decision on what action to take, based on the perceived state $s_t$. 
A policy can be:
\begin{itemize}
    \item Deterministic: $ a_t = \pi(s_t) $, i.e. a direct mapping between states and actions;
    \item Stochastic:  $ \pi(a_t|s_t) $, a probability distribution over actions, given their state.
\end{itemize}
DDPG \cite{DDPG} and TD3 \cite{td3} are examples of RL algorithms that learn a deterministic policy, while SAC \cite{sac} is a RL algorithm that learns a stochastic policy.

The standard mathematical formalism used to describe the agent environment is a \textit{Markov Decision Process (MDP)}. An MDP is a 5-tuple, $ \langle S,A,R,P,\rho_0  \rangle $, where : 
\begin{itemize}
    \item $ S $ is the set of all valid states,
    \item $A$ is the set of all valid actions,
    \item $ R : S \times A \rightarrow{} {\rm I\!R} $ is the reward function, with $r_t = R(s_t,a_t, s_{t+1})$, 
    \item $ P : S\times A \rightarrow{} P(s) $ is the transition probability function, with $ P(s_{t+1} | s_t, a_t ) $ being the probability of transitioning into state $ s_{t+1} $ starting from state $ s_t $ and taking action $a_t$,
    \item $\rho_0(s)$ is the starting state distribution.
\end{itemize}
Markov Decision Processes obey the \textit{Markov property}: a transition only depends on the most recent state and action (and not on states/actions that precede the most recent ones).

The goal in RL is to learn a policy $ \pi $ which maximizes the so-called \textit{expected return}, which can be computed as:
\begin{eqnarray*}
     R(\tau) &=& \sum_{t=0}^{\infty} \gamma^t r_t \hspace{1cm} 
\end{eqnarray*}

\noindent
A discount factor $ \gamma \in (0,1) $ is needed for convergence. It determines how much the agent cares about rewards in the distant future relative to those in the immediate future.
$\tau$ is a sequence of states and actions in the environment $\tau=(a_0,s_0, a_1, s_1 ...)$, named \textit{trajectory} or \textit{episode}.
Testing an Android app can be seen as a task divided into finite-length episodes.

To estimate the expected return in practice, it is convenient to express it using a \textit{value function} and an \textit{action-value function}. The value function $ V^\pi(s) $ is defined as the expected return starting in a state $s$ and then acting according to a given policy $ \pi $:

\vspace{-0.3cm}
\[V^\pi (s) = E[R_t| s_0 = s] \]

The action-value function $Q^{\pi}(s,a) $ can be used to describe the expected return after taking an action $ a $ in state $ s $ and thereafter following the policy $ \pi $:

\[ Q^{\pi}(s,a) = E[R_t|s,a] \]

Correspondingly, we can define the \textit{optimal value function}, $V^*(s)$, as the $ V^\pi(s) $ that gives the highest expected return when starting in state $ s $ and  acting according to the optimal policy in the environment. The \textit{optimal action-value function}, $Q^*(s,a)$,  gives the highest achievable expected return under the constraints that the process starts at state $s$, takes an action $a$ and then acts according to the optimal policy in the environment.

A policy that chooses greedy actions only with respect to $Q^*$ is optimal, i.e., knowledge of $Q^*$ alone is sufficient for finding the optimal policy. As a result, if we have $ Q^* $, we can directly obtain the optimal action, $ a^*(s) $, via $a^*(s) = \argmax_a Q^*(s,a)$.
The optimal value function $V^*(s)$ and action-value function $Q^*(s,a)$ can be computed by means of a recursive relationship known as the \textit{optimal Bellman equations}:

\vspace{-0.3cm}
\begin{eqnarray*} 
    V^*(s_t) &=& \max_a E[ r(s_t,a_t) + \gamma V^*(s_{t+1})] \\
    Q^*(s_t, a_t) &=& E[ r(s_t,a_t) + \gamma \max_{a_{t+1}} [Q^*(s_{t+1}, a_{t+1}) ] ]
\end{eqnarray*}

\subsection{Tabular RL}

Tabular techniques refer to RL algorithms where the state and action spaces are approximated using value functions defined by means of tables. 
In particular, \textit{Q-Learning} \cite{watkins1992q} is one of the most adopted algorithms of this family. 
Q-Learning aims at learning a so-called \textit{Q-Table}, i.e., a table that contains the value of each state-action pair. A Q-Table represents the current estimate of the action-value function $Q(s,a)$. Every time an action $a_t$ is taken and a state $s_t$ is reached, the associated state-action value in the Q-Table is updated as follows:

\vspace{-0.2cm}
\begin{equation*}
    Q(s_t, a_t):=Q(s_t,a_t) + \alpha(r_t + \gamma \max_a Q(s_{t+1}, a) - Q(s_t,a_t))
\end{equation*} 

\noindent
where $\alpha$ is the learning rate (between 0 and 1); $\gamma$ is the discount factor,  applied to the future reward. Typically, $\gamma$  ranges from 0.8 to 0.99 \cite{DQN} \cite{DDPG} \cite{td3}, while $\max_a Q(s_{t+1}, a)$ gives the maximum value for future rewards across all actions. It is used to update the reward for the current state-action pair.

RL algorithms based on the tabular approach do not scale to high-dimensional problems, because in such cases it is difficult to manually define a good initial Q-Table. In case a good initial Q-Table is not available, convergence to the optimal table by means of the update rule described above is too slow \cite{DDPG}.

\subsection{Deep Reinforcement Learning}

In large or unbounded discrete spaces, where representing all states and actions in a Q-Table is impractical, tabular methods become highly unstable and incapable to learn a successful policy\cite{DQN}. The rise of deep learning, relying on the powerful function approximation properties of deep neural networks, has provided new tools to overcome these limitations. One of the first deep reinforcement learning algorithms is DQN (Deep Q-Networks) \cite{DQN}. 

DQN uses convolutional neural networks to approximate the computation of the action-value function $Q^{\pi}$. Training of such neural networks is achieved by means of \textit{memory replay}: the last $N$ experience tuples are sampled uniformly and replayed when updating the weights of the networks.

While DQN can indeed solve problems with high-dimensional observation spaces, it can only handle discrete and low-dimensional action spaces.
The recent advancements over DQN described in the following paragraphs (namely, DDPG, TD3 and SAC) overcome such limitation and allow dealing with high-dimensional action spaces.

\subsubsection*{\textbf{Deep Deterministic Policy Gradient (DDPG)}}

DDPG \cite{DDPG} is an \textit{Actor-Critic} algorithm, i.e., it includes two roles: the \textit{Critic}, which estimates the value function, and the \textit{Actor}, which updates the policy $\pi$ in the direction suggested by the Critic.
It is based on a deterministic policy gradient \cite{DPG} that can operate over continuous action spaces. 

More specifically,
the Critic has the objective of learning an approximation of the function $ Q^*(s)$.
Suppose that the approximator is a neural network $ Q_{\phi}(s,a)$, with parameters $\phi$, and that we have a set \textit{D} of transitions $(s_t,a_t,r_t,s_{t+1},d)$, where \textit{d=1} if $s_{t+1}$ is a final state; 0 otherwise.
Such set \textit{D} of previous experiences is called \textit{replay buffer}. To avoid overfitting, the replay buffer should be large enough. 
The loss function of the neural approximator is the so-called \textit{Mean-Squared Bellman Error (MSBE)}, which measures how close $ Q_{\theta}(s,a)$ is to satisfying the Bellman equation:
\begin{equation*}
    L(\phi,D) = E[(Q_{\phi}(s_t,a_t) -  (r_t + \gamma(1-d)\max_{a_{t+1}}Q_{\phi}(s_{t+1},a_{t+1})))^2]
\end{equation*}

The term subtracted from $Q_{\phi}(s_t,a_t)$  is named the \textit{target}, because  minimization of MSBE makes the Q-function as close as possible to this value. 
Since the target depends recursively on the same parameter $\phi$ to train,  MSBE minimization can become unstable. The solution is to use a second neural network, called \textit{target network}, whose parameters are updated to be close to $ \phi $, but with some time delay that gives stability to the process. 

\noindent

The Actor's goal is to learn a deterministic policy $ \pi_\theta(s) $ which maximizes $ Q_\phi(s,a)$. Because the action space is continuous and we assume the Q-function is differentiable with respect to the action, we can use gradient ascent with respect to the policy parameter.

\subsubsection*{\textbf{Twin Delayed DDPG (TD3)}}
\label{TD3}

Although DDPG can often achieve good performance, it tends to be susceptible to critical tuning of its hyperparameters. This happens when the hyperparameters of the Q-function make it overestimate the Q-values, which in turn leads to policy failure, as the Actor is actually exploiting such errors as guidance.

TD3 \cite{td3} is an algorithm which addresses this issue by introducing three major changes, mostly on the Critic side:
(1) \textit{Clipped Double Q-Learning}: TD3 learns two Q-functions instead of one, and uses the smaller of the two Q-values as the target in the MSBE function. 
(2) \textit{Delayed Policy Update}: TD3 updates the policy and the target networks less frequently than the Q-function. 
(3) \textit{Target Policy Smoothing}: TD3 adds noise to the target action to make it harder for the policy to exploit Q-function errors, by smoothing out Q across changes of the action.

\subsubsection*{\textbf{Soft Actor Critic (SAC)}}

The central feature of SAC \cite{sac} is entropy regularization. Using the entropy-regularized method, an agent gets a bonus reward at each time step which is proportional to the entropy of the policy at that time step. In fact, differently from TD3, the policy of SAC is non deterministic and inclusion of entropy in the reward aims at promoting policies with a wider spread of alternatives to choose stochastically from.
The RL problem becomes the problem of finding the optimal policy $\pi^*$ according to the following equation:

\vspace{-0.3cm}
\begin{equation*}
    \pi^* = \argmax_{\pi} E[\sum_{t=0}^{\infty} \gamma^t(R(s_t,a_t,s_{t+1}) + \alpha H(\pi(\cdot|s_t))) ]
\end{equation*}

\noindent
where $H$ is the entropy, and $\alpha > 0$ is the trade-off coefficient.

\section{Approach}

\begin{figure}[htb]
    \centering
\includegraphics[width=80mm]{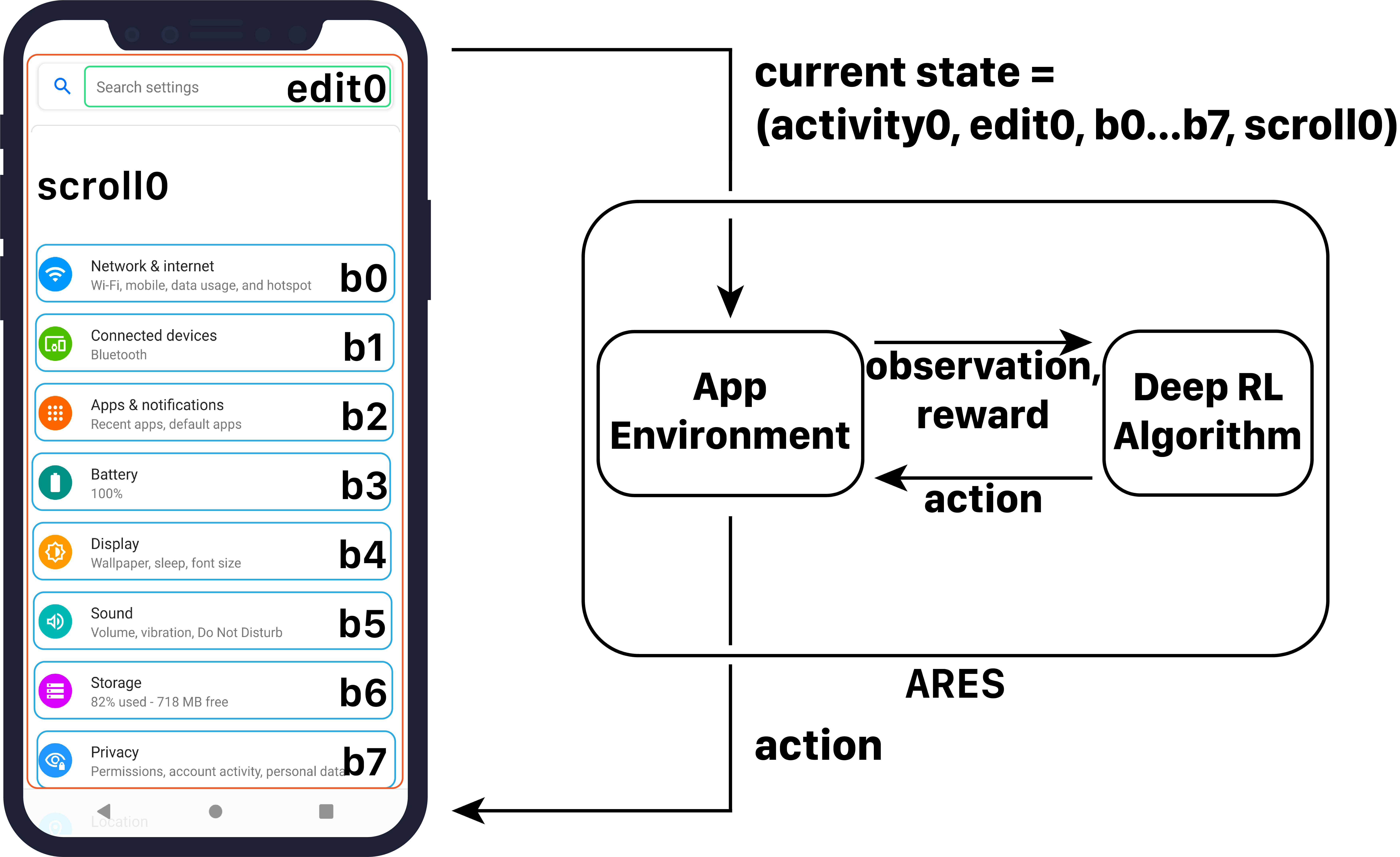}
    \caption{The \toolname{} testing workflow}
    \label{fig:workflow}
\end{figure}

In this section, we describe \toolname{} (Application of  REinforcement learning to android Software testing), our approach to black-box Android GUI testing based on Deep RL. The overview of the approach is shown in Figure \ref{fig:workflow}. 
The RL \textit{environment} is represented by the Android application under test (AUT), which is subject to several interaction steps. At each time step, assuming the GUI state is $s_{t}$,
\toolname{} first takes an action $a_t$. Then, it observes the new GUI state $s_{t+1}$ of the AUT and a reward $r_t$ is computed. 
Intuitively, if the new state $s_{t+1}$ is similar to the prior state $s_{t}$, the reward is negative. Otherwise, the reward is a large positive value. In this way, \toolname{} promotes the exploration of new states in the AUT, under the assumption that this is useful to test the application more thoroughly.

The reward is used to update the neural network, which learns how to guide the Deep RL algorithm to explore the AUT in depth. The actual update strategy depends on the deep RL algorithm being used (either DDPG, TD3 or SAC). 

\subsection{Problem Formulation}
\label{sub:ProblemFormulation}
To apply RL, we have to map the problem of Android black-box testing to the standard mathematical formalization of RL: an MDP, defined by the 5-tuple, $ \langle S,A,R,P,\rho_0 \rangle $. Moreover, we have to map the testing problem onto an RL task divided into several finite-length episodes.

\subsubsection*{\textbf{State Representation}}
Our approach is black-box, because it does not access the source code of the AUT. It only relies on the GUI of the AUT. 
The state $s_t \in S$ is defined as a combined state $(a_0, ... a_n, w_0, ... w_m)$. The first part of the state $a_0, ... a_n$ is a one hot encoding of the current activity, i.e., $a_i$ is equal to 1 only if the currently displayed activity is the $i$-th activity, it is equal to 0 for all the other activities. In the second part of the state vector, $w_j$ is equal to 1 if the $j$-th widget is available in the current activity; it is equal to 0 otherwise.

\subsubsection*{\textbf{Action Representation}}
User interaction events in the AUT are mapped to the action set $A$ of the MDP. \toolname{} infers executable events in the current state by analyzing the dumped widgets and their attributes (i.e., \textsl{clickable}, \textsl{long-clickable}, and \textsl{scrollable}). In addition to the widget-level actions we also model two system-level actions, namely \textsl{toggle internet connection} and \textsl{rotate screen}.
These system-level actions are the only system actions that can be easily tested. In fact, since Android version 7.0, testing other system-level actions (like those implemented in Stoat \cite{DBLP:conf/sigsoft/SuMCWYYPLS17}) would depend on the Android version used \cite{api19}, \cite{api25}, and would require a rooted device \cite{broadcasts}. In fact, working with a rooted device compromise the testing of protected apps \cite{root_apps}, such as apps in the finance category. 

Each action $a$ is 3-dimensional:
the first dimension represents the widget \toolname{} is going to interact with or the identifier of a system action. The second dimension specifies a string to be used as text input, if needed. Actually, an index pointing to an entry in a dictionary of predefined strings is used for this dimension. The third dimension depends on the context: when the selected widget is both \textit{clickable} and \textit{long-clickable}, the third action determines which of the two actions to take. When \toolname{} interacts with a \textit{scrollable} object, the third dimension determines the scrolling direction. 

\subsubsection*{\textbf{Transition Probability Function}}
The transition function $P$ determines which state the application can transit to after \toolname{} has taken an action. In our case, this is decided solely by the execution of the AUT: \toolname{}  observes the process passively, collecting the new state after the transition has taken place. 

\subsubsection*{\textbf{Reward Function}}
The RL algorithm used by \toolname{} receives a reward $r_t \in R$ every time it executes an action $a_t$. We define the following reward function:

\begin{equation} \label{eq:reward}
    r_t =
  \begin{cases}
    \Gamma_1       & \quad \text{if } act(s_{t+1}) \not \in act(E_t) \text{ or \textit{crash}}\\
    -\Gamma_2         & \quad \text{if } pack(act(s_{t+1})) \neq pack(AUT)\\
    -\Gamma_3   & \quad \text{otherwise}.
  \end{cases}
\end{equation}

\noindent
with $\Gamma_1 \gg \Gamma_2 \gg \Gamma_3$ (in our implementation $\Gamma_1 = 1000$, $\Gamma_2 = 100$, $\Gamma_3 = 1$).

The exploration of \toolname{} is divided into \textit{episodes}. At time $t$, the reward $r_t$ is high ($\Gamma_1$) if \toolname{} was able to transition to an activity never observed during the current episode $E_t$ (i.e., the next activity $act(s_{t+1})$ does not belong to the set of activities encountered so far in $E_t$): if a new episode is started at $t+1$, its set of activities is reset: $act(E_{t+1}) = \emptyset$.

Resetting the set of encountered activities  at the beginning of each new episode is a technique that encourages \toolname{} to visit and explore the highest number of activities in each episode, so as to continuously reinforce its explorative behaviors.
In contrast, giving a positive reward for each new activity only once, or, on the contrary, each time we see an activity change, would result respectively in too sparse, singular reinforcement feedback or in rewarding of cycling behaviors \cite{faulty_rewards}.

The reward is high ($\Gamma_1$) also when a faulty behavior (\textit{crash}) occurs.
It is very low ($-\Gamma_2$) when the displayed activity does not belong to the AUT (i.e., the package of the current activity, $pack(act(s_{t+1}))$,  is not the package of the AUT), as we aim to explore the current AUT only.
In all other cases, the reward is moderately negative ($-\Gamma_3$), as the exploration still remains inside the AUT, even if nothing new has been discovered.

\section{Implementation}

\toolname{} features a custom environment based on the OpenAI Gym \cite{gym} interface, which is a de-facto standard in the RL field. OpenAI Gym is a toolkit for designing and comparing RL algorithms and includes a number of built-in environments. It also includes guidelines for the definition of custom environments.
Our custom environment interacts with the Android emulator \cite{emulator} using the Appium Test Automator Framework \cite{Appium}.

\subsection{Tool Overview}

As soon as it is launched, \toolname{} leverages Appium to dump the widget hierarchy of the GUI in the starting activity of the AUT. The widget hierarchy is  analyzed searching for clickable, long-clickable and scrollable widgets. These widgets are afterwards stored in a dictionary containing several associated attributes (e.g., resource-id, clickable, long-clickable, scrollable, etc.) and compose the \textit{action vector}, i.e., the vector of executable actions in the current state.
At each time step, \toolname{} takes an action according to the behavior of the exploration algorithm. Once the action has been fully processed, \toolname{} extracts the new widget hierarchy from the current GUI and calculates its MD5 hash value. If it has the same MD5 value of the previous state, \toolname{} leaves the action vector unchanged. If the MD5 value does not match, \toolname{} updates the action vector.
\toolname{} performs the observation of the AUT state and returns the combined vector of activities and widgets.
\toolname{} organizes the testing of each app as a task divided into finite-length episodes. The goal of \toolname{} is to maximize the total reward received during each episode. Every episode lasts at least 250 time steps. Its duration is shorter only if the app crashes. Once an episode comes to an end, the app is restarted and \toolname{}  uses the acquired knowledge to explore the app more thoroughly in the next episode.

\subsection{Application Environment}

The application environment is responsible for handling the actions  to interact with the AUT. Since the environment follows the guidelines of the Gym interface, it is structured as a class with two key functions. The first function \verb|init(desired_capabilities)| is the initialization of the class. The additional parameter \verb|desired_capabilities| consists of a dictionary containing the emulator setup along with the application to be tested. The second function is the \verb|step(a)| function, that takes an action \verb|a| as command and returns a list of objects, including \verb|observation| (current AUT state) and \verb|reward|.



\subsection{Algorithm Implementation}

\toolname{} is a modular framework that adopts a plugin architecture for the integration of the RL algorithm to use. Hence, extension with a new exploration algorithm can be easily achieved.
In the current implementation, \toolname{} provides five different exploration strategies: (1) random, (2) Q-Learning, (3) DDPG, (4) SAC, (5) TD3.
The random algorithm interacts with the AUT by randomly selecting an action from those in the action vector.
Compared to Monkey \cite{Monkey}, our random approach performs better, since it selects only actions from the action vector. In fact, Monkey generates random, low level events on the whole GUI, which could target no actual widget and then be discarded.

Our Q-Learning strategy implements the algorithm by Watkins and Dayan \cite{watkins1992q}.
The deep RL algorithms available in \toolname{} are DDPG, SAC and TD3. Their implementation comes from the Python library \textit{Stable Baselines} \cite{stable-baselines}, and allow \toolname{} to save the status of the neural network as a policy file, at the end of the exploration. In this way, the policy can be loaded and reused on a new version of the AUT at a later stage, rather than restarting \toolname{} from scratch each time a new AUT version is released.
\toolname{} is publicly available as open source software at
\url{https://github.com/H2SO4T/ARES}.

\subsection{Compatibility}

\toolname{} has been successfully tested on Windows 10, MacOS 11.1 (and older), Ubuntu 20 (and older) and Scientific Linux 7.5. \toolname{} is fully compliant with parallel execution and enables parallel experiments to be performed on emulators or real devices, handling each instance in a completely separate manner. \toolname{} is also compatible with several Android versions (i.e., it has been successfully tested on Android 6.0, 7.0, 7.1, 8.0, 8.1, 9.0, and 10.0).
Moreover, since \toolname{} is based on the standard OpenAIGym, new algorithms and exploration strategies can be easily added to the tool.
\section{Fast Android Test Environment (FATE)}

Deep RL algorithms require fine tuning, which is expensive on real apps. Therefore, we developed FATE, a simulation environment for fast Android testing. FATE models only the navigation constraints of Android apps, so it can be used to efficiently compare alternative testing algorithms and to quickly tune their corresponding hyperparameters. After this algorithm selection and tuning phase through FATE is completed, the selected algorithms and their configurations are ported to \toolname{} to test real apps.

\subsection{Model-based Prototyping}
\label{sub:ImplProt}

In FATE, developers model an Android app by means of a deterministic Finite State Machine (FSM) $\mathcal{F} = (\Sigma, S, s_0, \delta, F)$, where $\Sigma$ is a set of events, $S$ a set of states with $s_0$ the initial state and $F$ the set of final states, and $\delta$ the state transition function $\delta: S \times \Sigma \longrightarrow 2^S$.
The \textit{states} $S$ of the FSM correspond to the activities of the app, while the \textit{events} $\Sigma$ trigger the transitions between activities, which in turn are modeled as a transition table $\delta$.  Events represent the clickable widgets (e.g., buttons) available in each activity. Transitions have access to a set of \textit{global variables} and possess, among others, the following attributes: \textit{ID}, \textit{type}, \textit{active} (boolean attribute), \textit{guard} (boolean expression that prevents the transition from being taken if it evaluates to false), \textit{set} (new values to be assigned to global variables), \textit{destination} (target activity, i.e., value of $\delta$).

\begin{figure*}[htb]
    \centering
\includegraphics[width=135mm]{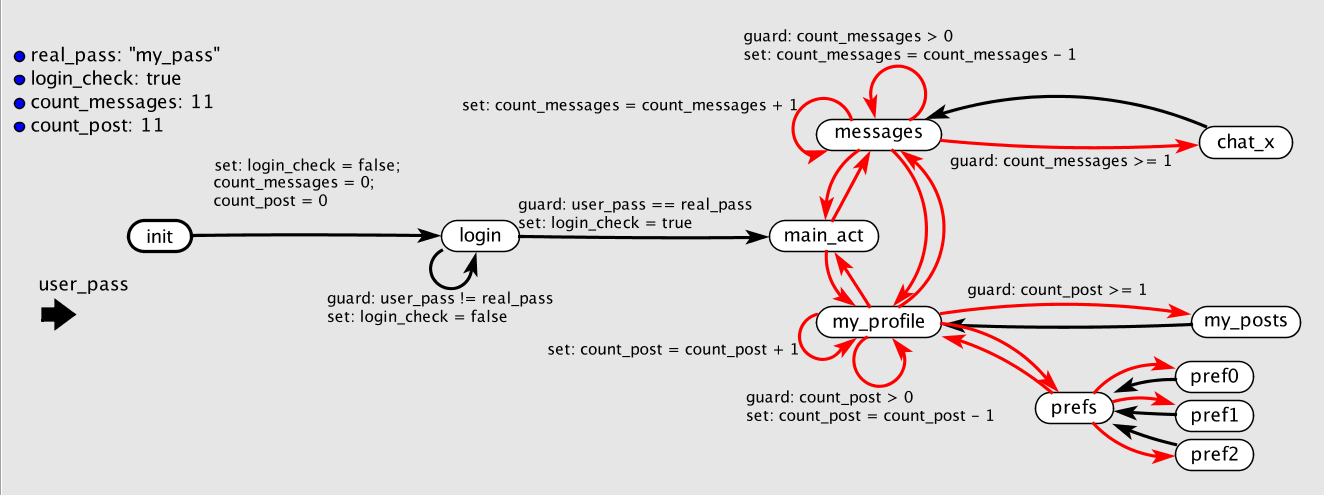}
    \caption{FATE model of  \textit{Social Network}: global variables are shown on the top-left; inputs on the bottom-left; red edges indicate non deterministic transitions.}
    \label{fig:FSM}
\end{figure*}

Figure \ref{fig:FSM} shows the FATE model of the prototypical app \textit{Social Network}. 
To build such model, developers can  use Ptolemy  \cite{ptolemy} to graphically draw a FSM that mimics the behavior of the application. While creating a FSM with Ptolemy is not mandatory in FATE, it simplifies the job of designing a logically correct model, thanks to the checks it performs.
Then, FATE translates automatically the Ptolemy model (saved in XML format) into a Json file that replicates the structure and behavior of the Ptolemy model. The Json model translation has two main fields: \verb|global_vars| and \verb|nodes|. The first contains a list of global variables organized by \verb|name| and \verb|value|. The latter contains a list of all the activities. Each activity is characterized by a \verb|node_id| and a list of corresponding node transitions, each including all the respective transition attributes. 

The model of \textit{Social Network} in Figure \ref{fig:FSM} contains a transition from \verb|login| to \verb|main_act| which is subjected to the guard \verb|user_pass == real_pass|, i.e., the entered password must be correct in order for the transition to be taken. In the Json model, such transition is coded as:
\begin{lstlisting}[language=Python, style=mystyle]
"transition": {
    "transition_id": 0,
    "type": "button",
    "active": true,
    "guard": "user_pass == real_pass",
    "set": null,
    "destination": "main_act"
}
\end{lstlisting}
Another example of guarded transition is the one between \verb|messages| and \verb|chat_x|. The guard \verb|count_messages >= 1| checks whether there exists at least one message from which a chat thread can be started. In the Json model this is coded as:
\newline
\begin{lstlisting}[language=Python, style=mystyle]
"transition": {
    "transition_id": 0,
    "type": "button",
    "active": true,
    "guard": "count_messages >= 1",
    "set": null,
    "destination": "main_act"
}
\end{lstlisting}

In FATE, a Python environment compliant with the OpenAI Gym standard takes as input the Json app model and tests it automatically using the selected algorithm. The available algorithms  are: (1) Random, (2) Q-Learning, (3) DDPG, (4) SAC, (5) TD3. FATE was built with modularity in mind and new exploration algorithms can be easily added to the tool.

Compared to testing an Android app through Espresso or Appium, FATE makes test case execution extremely faster, because there is no need to interact with the app via its GUI. Moreover, the application navigation logic is simulated by the transition function $\delta$, which makes it usually  much faster to execute. 
As a consequence, developers can run a large number of experiments, evaluate multiple algorithms, check various algorithm  or  application configurations, and find the optimal set of hyperparameters, all of which would be prohibitively expensive to execute on a standard Android testing platform.
FATE is publicly available as open source software at \url{https://github.com/H2SO4T/ARES}.

\subsection{Representative Family of Models}
\label{models}

For fast evaluation of the Deep RL algorithms implemented in \toolname{}, we modeled four Android apps using FATE. Each model is configurable with a variable degree of complexity.
To obtain a set of app models that are representative of the most common apps used in everyday life, we inspected AppBrain (a website that aggregates Google Play Store statistics and rankings) \cite{appbrain} and selected four different and representative categories from the top-ten: Music \& Audio,  Lifestyle, Business, and Shopping. From each category we then selected and modeled in FATE one prototypical app: \textit{Player}, \textit{Social Network}, \textit{Bank} and \textit{Market Place}.

The simplest scenario is \textit{Player}. It features a wide number of activities arranged in a tree-like structure. It reflects the generalization of a variety of applications, including  apps or app components to manage the settings and to stream/add/remove media contents.
\textit{Social Network} (see Figure \ref{fig:FSM}) starts by prompting a login activity with fields for username and password. Following the login activity we have several activities that replicate the behavior of a standard social network, including a basic chat activity.
The \textit{Bank} model is characterized by the presence of inner password-protected operations. 
\textit{Market Place} models a typical app for e-commerce: the user can search for goods, login, purchase products, and monitor the orders.
The four representative app models used in this work are publicly available inside the FATE tool.
\section{Evaluation}

We seek to address the following research questions, split between the following two studies:

\textbf{Study 1 (FATE)}:
\begin{itemize}
    \item \textbf{RQ1} \textit{Which deep RL algorithm and which algorithm configuration performs better on the synthetic apps?} 
    \item \textbf{RQ2} \textit{How does activity coverage vary as the model of the AUT becomes increasingly difficult to explore?} 
    \item \textbf{RQ3} \textit{What are the features of the synthetic apps that allow deep RL to perform better than Q-Learning?}
    \item \textbf{RQ4} \textit{Are the results of synthetic apps comparable to those of the their translated counterparts?}
\end{itemize}

In Study 1, we take advantage of the fast execution granted by FATE to compare alternative RL algorithms (RQ1) and determine their optimal configuration (see Appendix 1). Since in this study we use synthetic apps generated from models (i.e., Player, Social Network, Bank and Market Place), coverage is measured at the  granularity of Android activities. In fact, there is no source code implementing the business logic. 
Text inputs are chosen from a pool of 20 strings, which include credentials necessary to pass through the login activities.
To account for  non determinism, we executed each algorithm 60 times for each hyperparameter configuration of the algorithms, and applied the Wilcoxon non-parametric statistical test to draw conclusions on the difference between algorithms and configurations, adopting the conventional $p$-value threshold at alpha. Since multiple pairwise comparisons are performed with overlapping data, the chance to reject true null hypotheses may increase (type I error). To control this problem, we adopt the Holm-Bonferroni correction~\cite{Holm}, that consists in using more strict significance levels when the number of pairwise tests increases. 

Each run has a length of 4000 time steps, which is close to an hour of testing in a real Android test setting. With FATE, 4000 times steps are executed approximately in 200 seconds.

To answer RQ2 we consider the best performing configuration of the Deep RL algorithms, as selected from RQ1, and gradually increase the exploration complexity of the apps. Specifically, \textit{20\_strings, 40\_strings, 80\_strings}  indicate an increasing size of the string pool (from 20, to 40 and 80), while  \textit{augmented\_5} and \textit{augmented\_10}  indicate an increasing size of the self navigation links (with 5 or 10 ``dummy'' buttons) within the login activities. 
For the assessment, we adopt the widely used metric AUC (Area Under the Curve), measuring the area below the activity coverage plot over time.
To account for the non determinism of the algorithms, we repeated each experiment 30 times and applied the Wilcoxon non-parametric statistical test.

In RQ3 we investigate qualitatively the cases where Deep RL is superior to tabular Q-Learning. With RQ4, we want to understand if results obtained on synthetic app models run by FATE correlate with those obtained when executing the same apps, once they are translated into real Android apps in Java code that can be executed by \toolname{}. 
In particular, we translated 3 synthetic apps to Java/Android: Social Network, Bank and Market Place. We then repeated the scenario \textit{40\_strings} on them using \toolname{}. We compare the rankings of the algorithms being evaluated that are produced respectively by FATE and by \toolname{}.

\textbf{Study 2 (\toolname{})}:
\begin{itemize}
    \item\textbf{RQ5} \textit{How do code coverage and time-dependent code coverage compare between Random, Q-Learning, DDPG, and SAC?}
    \item\textbf{RQ6} \textit{What are the fault exposure capabilities of the alternative approaches?}
    \item\textbf{RQ7} \textit{What features of the real apps make deep RL perform better than Q-Learning?}
    \item\textbf{RQ8} \textit{How does ARES compare with state-of-the-art tools in terms of coverage and bug detection?}
\end{itemize}

In Study 2, we use real apps and compare the alternative Deep RL algorithms between each other and with Random and tabular Q-Learning. At last, we compare ARES to state of the art testing tools.

To address RQ5-RQ6 and RQ7, we randomly selected 41 apps among the 500 most starred F-Droid apps available on GitHub.
We consider coverage at the source code level and compare both the final coverage as well as the coverage increase over time (RQ5). 
To obtain coverage data we instrumented each app by means of JaCoCo \cite{jacoco}.
As in Study 1, we measured AUC and compared AUC values using the Wilcoxon statistical test with significance level set to 0.05 (with correction). 
We exclude TD3 from the comparison, since it performed consistently worse than the other RL algorithms on synthetic apps. 

In addition to code coverage, we also report the number of failures (unique app crashes) triggered by each approach (RQ6).
To measure the number of unique crashes observed, we parsed the output of Logcat and (1) removed all crashes that do not contain the package name of the app; (2) extracted the stack trace; (3) computed the hash code of the sanitized stack trace, to uniquely identify it.
With RQ7 we analyze qualitatively the different performance of Deep RL vs Q-Learning on real apps. 

To address RQ8, we evaluate and compare \toolname{} and state of the art tools in terms of code coverage and number of crashes, using two different sets of apps under test, RQ8-a and RQ8-b, that accommodate the different requirements and constraints of the tools being compared. As state of the art tools, we selected Monkey \cite{Monkey}, Sapienz \cite{mao2016sapienz}, TimeMachine \cite{dong2020timetravel} and Q-Testing \cite{qtest}.
In RQ8-a we compare ARES, Monkey, Sapienz and TimeMachine on a set of 68 apps coming from AndroTest \cite{androtest}. These apps are instrumented using Emma \cite{emma}, the same coverage tool that is used in Sapienz and TimeMachine.
In RQ8-b, we compare ARES to Q-Testing on a set of ten apps instrumented using JaCoCo, the coverage tool supported by Q-Testing.

All experimental data was generated and processed automatically. Each experiment was conducted with a 1 hour timeout and was repeated 10 times, for a total of 4560 hours ($\approx$ 190 days). The emulators involved in the study are equipped with 2 GB of RAM and Android 10.0 (API Level 29) or Android 4.4 only for the tool comparison.

\subsection{Experimental Results: Study 1}

\begin{figure}[!h]
    \centering
    \includegraphics[width=0.8\textwidth]{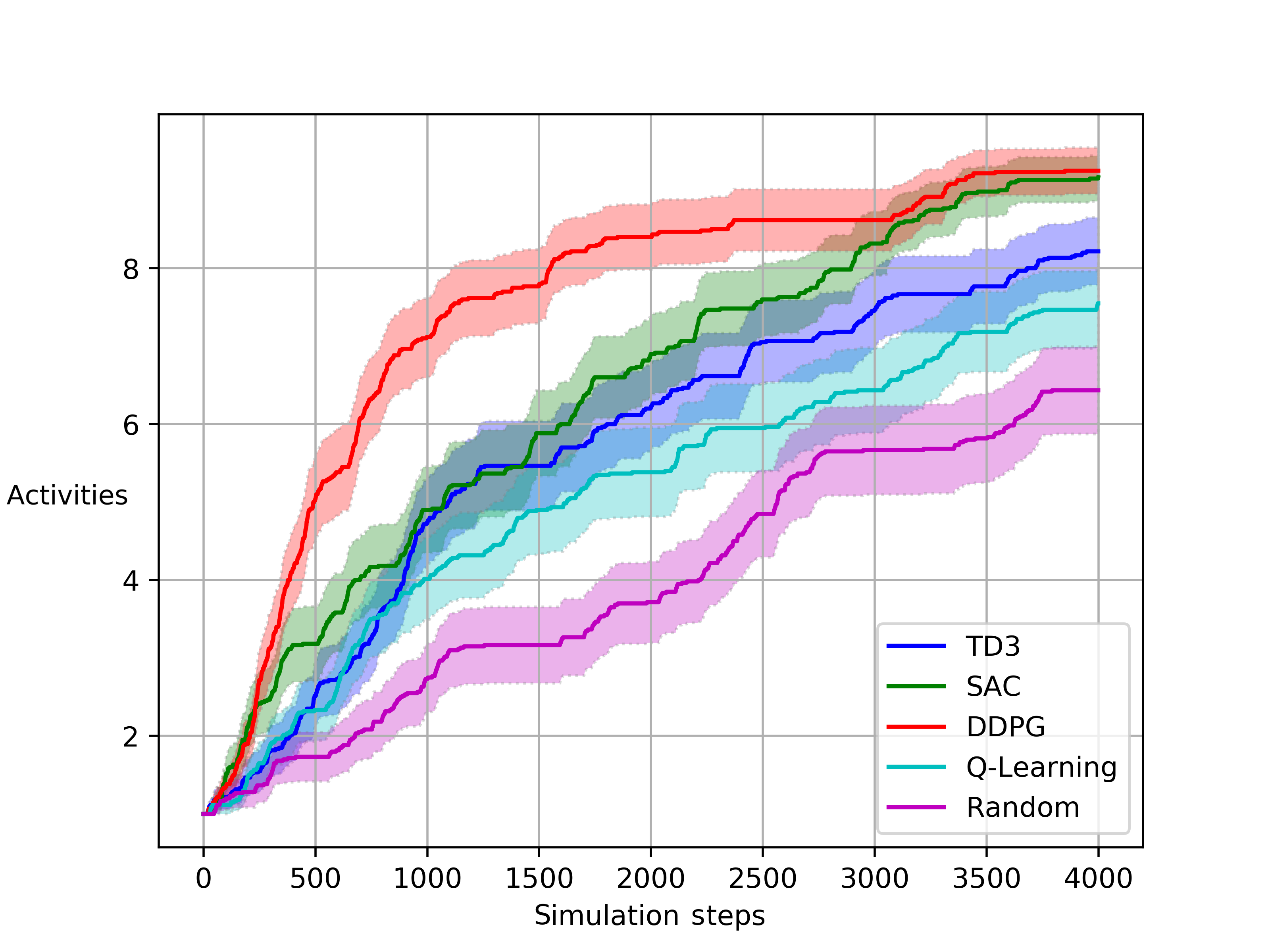}
    \caption{Activity coverage of the Social synthetic app in FATE}
    \label{fig:social_sem}
\end{figure}

\begin{table}[h]
\scriptsize
\centering
 \begin{tabularx}{0.67\textwidth}{c|c|c|c|c|c|c|c} 
 \hline
 \textbf{App} & \textbf{Config} & \textbf{Rand} & \textbf{Q-Learn} & \textbf{TD3} & \textbf{SAC} & \textbf{DDPG} & \textbf{Effect Size}\\
 \hline
 Player & \textit{20\_str} & 88719 & 89022 & 89903 & 89943 & \cellcolor[HTML]{C0C0C0}90337& - \\\hline
 \multirow{5}{*}{Social}  
 & \textit{20\_str} & 15840 & 20387 & 22809 & 25463 & \cellcolor[HTML]{C0C0C0}30008 & L(Rand), M(Q) \\
 & \textit{40\_str} & 9291 & 7737 & \cellcolor[HTML]{C0C0C0}14451 & 10361 & 7363 & S(DDPG) \\
 & \textit{80\_str} & 4535 & 5640 & 5730 & \cellcolor[HTML]{C0C0C0}7254 & 4774 & - \\
 & \textit{aug\_5} & 13960 & 15400 & 13094 & \cellcolor[HTML]{C0C0C0}17402 & 13385 & - \\
 & \textit{aug\_10} & 5291 & 3998 & \cellcolor[HTML]{C0C0C0}13737 & 11559 & 8870 & M(Q, Rand)\\\hline
 \multirow{5}{*}{Bank} 
 & \textit{20\_str} &  22894 & 21622 & 29159 & 28016 & \cellcolor[HTML]{C0C0C0}36977 & M(Q, Rand)\\
 & \textit{40\_str} &  9746 & 7750 & 9305 & \cellcolor[HTML]{C0C0C0}16535 & 6458 & S(Q, DDPG)\\
 & \textit{80\_str} &  3998 & 4843 & 4776 & \cellcolor[HTML]{C0C0C0}5621 & 4798 & - \\
 & \textit{aug\_5} & 12815 & 8634 & 8702 & \cellcolor[HTML]{C0C0C0}14914 & 11472 & - \\
 & \textit{aug\_10} & 4121 & 6289 & 13289 & 14195 & \cellcolor[HTML]{C0C0C0}15361 & M(Q, Rand)\\\hline
 \multirow{5}{*}{Market}
 & \textit{20\_str} &  19236 & 18471 & 20980 & 23403 & \cellcolor[HTML]{C0C0C0}25923 & - \\
 & \textit{40\_str} &  15943 & \cellcolor[HTML]{C0C0C0}16496 & 15949 & 15936 & 16318 & - \\
 & \textit{80\_str} &  15944 & \cellcolor[HTML]{C0C0C0}15945 & 15935 & 15937 & 15932 & - \\
 & \textit{aug\_5} & 18917 & 16377 & 16500 & \cellcolor[HTML]{C0C0C0}21208 & 16027 & - \\
 & \textit{aug\_10} & 4121 & 6289 & 13289 & 14195 & \cellcolor[HTML]{C0C0C0}15361 & -\\
 \hline
 \end{tabularx}
 \vspace{3mm}
 \caption{AUC for synthetic apps: effect size between winner (shaded cell) and other algorithms is reported only when $p$-value is statistically significant (S = Small; M = Medium; L = Large)}
\label{table:synth_summary}
\end{table}

Figure \ref{fig:social_sem} shows the coverage growth for the  synthetic app Social. Each curve shows the mean over 60 runs. The shaded area around the mean represents the Standard Error of the Mean (\textit{SEM} = $\sigma / \sqrt{n}$, where $\sigma$ is the standard deviation and $n = 60$ the number of points). The highest activity coverage is obtained consistently by deep RL algorithms, which correspondingly have also higher AUC values. Table \ref{table:synth_summary} reports the AUC  obtained on the synthetic apps in all  tested configurations. Table \ref{table:synth_summary} also shows the Vargha-Delaney effect size in the case of a $p$-value $<$ $\alpha$ (with correction), between the winner algorithm (highest AUC) and the remainders. 

Results show that deep RL algorithms achieve higher coverage in most experiments. DDPG performs better in the simplest configuration, \textit{20\_strings}, while SAC performs better in almost all other configurations, including the most complex ones. Q-Learning prevails in only two scenarios belonging to \textit{Market Place}, but the difference from the other algorithms is not statistically significant ($p$-value $>$ $\alpha$).

\begin{tcolorbox}[,colback=gray!5,colframe=blue!40, boxrule=0.3mm]
RQ1: 
Results lead us to select DDPG and SAC as best performing deep RL algorithms, to be involved in Study 2 experiments.
\end{tcolorbox}
DDPG is selected due to its high performance in fairly simple scenarios;
SAC because of its ability to adapt and maintain good performance in the majority of scenarios.

\begin{tcolorbox}[,colback=gray!5,colframe=blue!40, boxrule=0.3mm]
RQ2: 
While in simple situations (e.g., the Player app), all algorithms achieve a high level of coverage, 
when things get more complex (e.g., when string pool increases), Deep RL algorithms retain higher coverage than all other algorithms.
\end{tcolorbox}

We have manually inspected the step by step exploration performed by Q-Learning and by the Deep RL algorithms. We found that login activities complicate substantially the exploration performed by Q-Learning. In fact, it is more difficult to reproduce the right username-password combination for a tabular Q-Learning algorithm, which has limited adaptation capabilities, whereas Deep RL algorithms memorize the right combination in the DNN used to guide the exploration.
In addition, large action spaces make it challenging for Q-Learning to learn an effective exploration strategy, while the DNNs used by deep RL algorithms can easily cope with large spaces of alternatives to choose from. This is confirmed by the performance degradation of Q-Learning as the string pool becomes larger or as new interactive elements (``dummy'' buttons) are added, which confuse Q-Learning during its exploration.

\begin{tcolorbox}[,colback=gray!5,colframe=blue!40, boxrule=0.3mm]

RQ3: The performance of Q-Learning declines in the presence of blocking activities that require specific input combinations that must be learned from past interactions or when the input/action space becomes excessively large, while Deep RL can learn how to cope with such obstacles thanks to the DNN employed to learn the exploration strategy.
\end{tcolorbox}

Table \ref{table:real_vs_sim} shows the ranking of the algorithms produced by \toolname{} vs FATE on the 3 apps that were translated from the synthetic FATE models 
to Java/Android. Below the ranking, Table \ref{table:real_vs_sim} shows the AUC values obtained by the respective algorithms.
The behaviors of the considered algorithms on synthetic (FATE) vs translated (\toolname{}) apps are very similar. The AUC values are quite close and the Spearman's correlation between AUC values across algorithms is 0.99 for Social, 0.89 for Bank and 0.99 for Market; it is 0.95 overall. All correlations are statistically significant at level 0.05.
\toolname{} required 450 hours to complete the experiments. FATE required around 10 hours, reducing the computation time by a factor 45, while producing similar results as \toolname{}.

\begin{table}[h]
\scriptsize
\centering
\begin{tabular}{cc||ccccc}
\hline
\textbf{App} & \textbf{Tool} & \multicolumn{4}{c}{\textbf{Ranking / AUC}} \\
\hline
Social & \toolname{} & Q-Learn & DDPG  & Rand  & SAC   & TD3   \\
                     & & 4: 7788    & 5: 6802  & 3: 9547 & 2: 9594 & 1: 15101 \\ 
                   
 & FATE  & Q-Learn & DDPG & Rand & SAC   & TD3   \\
 &       & 4:7737    & 5:7363 & 3:9291 & 2:10361 & 1:14451 \\

\hline
Bank & \toolname{} & Q-Learn & DDPG & Rand  & SAC   & TD3   \\
     &             & 4: 8614 & 5: 7976 & 3: 9344  & 1: 12138 & 2: 10932  \\ 
                
     & FATE & Q-Learn & DDPG  & Rand  & SAC   & TD3  \\
     &      & 4: 7750   & 5: 6458  & 2: 9746  & 1: 16535 & 3: 9305 \\
\hline
Market & \toolname{} & Q-Learn & DDPG  & Rand  & SAC   & TD3    \\
                   & & 1: 16866   & 2: 16788 & 4: 15936 & 5: 15930 & 3: 15944  \\ 
 & FATE & Q-Learn   & DDPG       & Rand   & SAC    & TD3   \\
 &      & 1: 16496  & 2: 16318   & 4: 15943  & 5: 15936  & 3: 15949 \\
\hline
\end{tabular}
\vspace{3mm}
 \caption{Ranking of algorithms produced by \toolname{} vs FATE; AUC values  below ranked algorithms}
\label{table:real_vs_sim}
\end{table}

\begin{tcolorbox}[,colback=gray!5,colframe=blue!40, boxrule=0.3mm]
RQ4: The results obtained on synthetic apps are comparable to those obtained on their translated counterparts. 
\end{tcolorbox}

\subsection{Experimental Results: Study 2}

\begin{table}[h]
\scriptsize
\begin{tabularx}{0.68\textwidth}{|l|>{\centering\arraybackslash}m{0.5cm}|>{\centering\arraybackslash}m{0.3cm}>{\centering\arraybackslash}m{0.4cm}>{\centering\arraybackslash}m{0.4cm}>{\centering\arraybackslash}m{0.6cm}|>{\centering\arraybackslash}m{0.3cm}>{\centering\arraybackslash}m{0.4cm}>{\centering\arraybackslash}m{0.4cm}>{\centering\arraybackslash}m{0.6cm}|}
\hline
\textbf{Applications} & \textbf{ELOC} &\textbf{Rand} & \textbf{Q} & \textbf{SAC} & \textbf{DDPG} & \textbf{Rand} & \textbf{Q} & \textbf{SAC} & \textbf{DDPG} \\
& & & \%Coverage(mean) & & & &\#Crashes(mean)& & \\
\hline
\hline
Silent-ping-sms & 263 & 41 & 41 & 41 & 41 & 0 & 0 & 0 & 0 \\
Drawablenotepad & 452 & 20 & 21 & {\cellcolor[rgb]{0.753,0.753,0.753}}26 & 25 & {\cellcolor[rgb]{0.753,0.753,0.753}}0.7 & 0.6 & {\cellcolor[rgb]{0.753,0.753,0.753}}0.7 & 0.1 \\
SmsMatrix & 466 & 23 & 20 & {\cellcolor[rgb]{0.753,0.753,0.753}}24 & 22 & {\cellcolor[rgb]{0.753,0.753,0.753}}1.2 & 0 & {\cellcolor[rgb]{0.753,0.753,0.753}}1.2 & 0.9 \\
Busybox & 540 & 75 & 73 & 74 & {\cellcolor[rgb]{0.753,0.753,0.753}}76 & 0 & 0 & 0 & 0 \\
WiFiKeyShare & 627 & {\cellcolor[rgb]{0.753,0.753,0.753}}37 & 36 & {\cellcolor[rgb]{0.753,0.753,0.753}}37 & {\cellcolor[rgb]{0.753,0.753,0.753}}37 & 0 & 0 & 0 & 0 \\
Talalarmo & 1094 & 69 & {\cellcolor[rgb]{0.753,0.753,0.753}}71 & {\cellcolor[rgb]{0.753,0.753,0.753}}71 & {\cellcolor[rgb]{0.753,0.753,0.753}}71 & 0 & \cellcolor[HTML]{C0C0C0}0.5 & \cellcolor[HTML]{C0C0C0}0.5 & 0 \\
AquaDroid & 1157 & 55 & 55 & 55 & 55 & {\cellcolor[rgb]{0.753,0.753,0.753}} 1.0 & 0.4 & 0.8 & 0.3 \\
Lexica & 1215 & 72 & 72 & 74 & {\cellcolor[rgb]{0.753,0.753,0.753}}75 & 0.3 & 0.1 & {\cellcolor[rgb]{0.753,0.753,0.753}}1.5 & \cellcolor[HTML]{C0C0C0} 1.2 \\
Loyalty-card-locker & 1228 & 41 & 37 & {\cellcolor[rgb]{0.753,0.753,0.753}}50 & 41 & 0.5 & 0.4 & {\cellcolor[rgb]{0.753,0.753,0.753}}0.8 & 0.1 \\
Dns66 & 1264 & 58 & 58 & 58 & 58 & 0.1 & 0 & 0 & \cellcolor[HTML]{C0C0C0}0.2 \\
Gpstest & 1311 & {\cellcolor[rgb]{0.753,0.753,0.753}}47 & 46 & {\cellcolor[rgb]{0.753,0.753,0.753}}47 & 46 & 0 & 0 & 0 & 0 \\
Memento & 1336 & {\cellcolor[rgb]{0.753,0.753,0.753}}77 & 76 & 74 & {\cellcolor[rgb]{0.753,0.753,0.753}}77 & 0 & 0 & 0 & 0 \\
Editor & 1547 & 50 & 46 & {\cellcolor[rgb]{0.753,0.753,0.753}}51 & 50 & 0 & 0 & 0 & 0 \\
AndOTP & 1560 & 20 & 25 & {\cellcolor[rgb]{0.753,0.753,0.753}}27 & 20 & 0.5 & 0.5 & {\cellcolor[rgb]{0.753,0.753,0.753}}0.7 & 0.2 \\
BookyMcBookface & 1595 & {\cellcolor[rgb]{0.753,0.753,0.753}}26 & 25 & 25 & 24 & 0 & 0 & 0 & 0 \\
Tuner & 2207 & {\cellcolor[rgb]{0.753,0.753,0.753}}80 & 74 & 79 & 75 & 0 & 0 & 0 & 0 \\
WifiAnalyzer & 2511 & 78 & 75 & {\cellcolor[rgb]{0.753,0.753,0.753}}80 & 79 & 0 & 0 & 0 & 0 \\
AdAway & 3064 & 38 & 37 & {\cellcolor[rgb]{0.753,0.753,0.753}}45 & 40 & 0 & 0 & \cellcolor[HTML]{C0C0C0}0.1 & \cellcolor[HTML]{C0C0C0}0.1 \\
Gpslogger & 3201 & {\cellcolor[rgb]{0.753,0.753,0.753}}36 & 31 & 32 & 28 & 0 & 0 & 0 & \cellcolor[HTML]{C0C0C0}0.1 \\
Connectbot & 3904 & 26 & 25 & {\cellcolor[rgb]{0.753,0.753,0.753}}28 & 18 & 0 & 0 & 0 & 0 \\
Neurolab & 3954 & {\cellcolor[rgb]{0.753,0.753,0.753}}29 & 28 & {\cellcolor[rgb]{0.753,0.753,0.753}}29 & 28 & 0 & 0.4 & 0.3 & {\cellcolor[rgb]{0.753,0.753,0.753}} 0.6 \\
Anuto & 4325 & 46 & 46 & {\cellcolor[rgb]{0.753,0.753,0.753}}47 & \cellcolor[HTML]{C0C0C0}47 & 0 & 0 & 0 & 0 \\
PassAndroid & 4569 & 1 & 1 & 1 & 1 & 0 & 0 & 0 & 0 \\
Markor & 4607 & 51 & 43 & {\cellcolor[rgb]{0.753,0.753,0.753}}53 & 41 & 0.3 & 0 & {\cellcolor[rgb]{0.753,0.753,0.753}}0.4 & 0 \\
Vanilla & 4747 & 29 & 34 & {\cellcolor[rgb]{0.753,0.753,0.753}}41 & 33 & 0 & 0 & 0 & 0 \\
\hline
\textbf{Average} & & 45 & 43.84 & 46.76 & 44.32 & 0.15 & 0.12 & 0.28 & 0.15 \\
\hline
\hline
Afwall & 5130 & 12 & 12 & {\cellcolor[rgb]{0.753,0.753,0.753}}16 & 13 & 0 & 0 & 0 & 0 \\
OpenTracks & 5260 & {\cellcolor[rgb]{0.753,0.753,0.753}}45 & 42 & 44 & {\cellcolor[rgb]{0.753,0.753,0.753}}45 & 0 & 0 & 0 & 0 \\
Opentasks & 5772 & 43 & 50 & {\cellcolor[rgb]{0.753,0.753,0.753}}53 & 44 & 0 & 0 & {\cellcolor[rgb]{0.753,0.753,0.753}}0.2 & 0 \\
UserLAnd & 5901 & 60 & 60 & 60 & 60 & 0.1 & 0.2 & {\cellcolor[rgb]{0.753,0.753,0.753}}0.4 & 0.2 \\
Simple-Solitaire & 5907 & 10 & 30 & {\cellcolor[rgb]{0.753,0.753,0.753}}31 & {\cellcolor[rgb]{0.753,0.753,0.753}}31 & 0 & {\cellcolor[rgb]{0.753,0.753,0.753}}0.4 & {\cellcolor[rgb]{0.753,0.753,0.753}}0.4 & 0.2 \\
Authorizer & 5923 & 5 & 5 & 5 & 5 & 0 & 0 & 0 & 0 \\
YalpStore & 6734 & 35 & 34 & {\cellcolor[rgb]{0.753,0.753,0.753}}38 & 33 & 0 & 0 & 0 & 0 \\
CameraRoll & 6836 & {\cellcolor[rgb]{0.753,0.753,0.753}}32 & 31 & 31 & {\cellcolor[rgb]{0.753,0.753,0.753}}32 & 0.8 & 0.1 & {\cellcolor[rgb]{0.753,0.753,0.753}}1.6 & 0.1 \\
AntennaPod & 7975 & 46 & 40 & {\cellcolor[rgb]{0.753,0.753,0.753}} 48 & 38 & 0.5 & 0.1 & {\cellcolor[rgb]{0.753,0.753,0.753}}0.8 & 0.4 \\
Phonograph & 8758 & 16 & 16 & 16 & 16 & 0 & 0 & 0 & 0 \\
\hline
\textbf{Average} & & 30.4 & 30.5 & 34.2 & 31.7 & 0.14 & 0.08 & 0.34 & 0.09  \\
\hline
\hline
MicroMathematics & 10506 & 35 & 35 & {\cellcolor[rgb]{0.753,0.753,0.753}}47 & 41 & 0 & 0 & 0 & 0 \\
LightningBrowser & 11961 & 35 & 36 & {\cellcolor[rgb]{0.753,0.753,0.753}}43 & 37 & 0 & 0 & {\cellcolor[rgb]{0.753,0.753,0.753}} 0.4 & 0.1 \\
Firefox-focus & 12482 & 33 & 34 & {\cellcolor[rgb]{0.753,0.753,0.753}}41 & 35 & 0.5 & 0.3 & {\cellcolor[rgb]{0.753,0.753,0.753}}0.8 & 0.1 \\ 
RedReader & 12958 & 42 & 42 & 44 & {\cellcolor[rgb]{0.753,0.753,0.753}}46 & 0 & 0 & \cellcolor[HTML]{C0C0C0}0.1 & 0 \\
Wikipedia & 23543 & 42 & 43 & {\cellcolor[rgb]{0.753,0.753,0.753}}44 & 41 & 0 & 0 & 0 & 0 \\
Slide & 30483 & {\cellcolor[rgb]{0.753,0.753,0.753}}19 & 17 & 18 & {\cellcolor[rgb]{0.753,0.753,0.753}}19 & 0.8 & 0.3 & \cellcolor[HTML]{C0C0C0}1.2 & 0.3 \\
\hline
\textbf{Average} & & 34.33 & 34.50 & 39.5 & 36.5 & 0.21 & 0.1 & 0.38 & 0.1 \\
\hline
\hline
\textbf{Total Average} & & 39.62 & 39.58 & 42.61 & 40.09 & 0.17 & 0.1 & 0.3 & 0.12 \\
\textbf{Unique crashes} & &  &  &  &  & 73 & 43 & 102 & 52 \\
\hline
\end{tabularx}\vspace{3mm}
\caption{Average coverage and number of crashes observed on 41 real open-source apps in 10 runs of \toolname{}}
\label{table:results}
\end{table}

Table \ref{table:results} shows  coverage and  crashes produced by each algorithm deployed in \toolname{}. The highest average coverage and average number of  crashes over 10 runs are shaded in gray for each app. We grouped the apps into three different size categories (Low-Medium, Medium, and High), depending on their ELOC (Executable Lines Of Code).
Inspecting the three categories, results show that the Deep RL algorithms arise more often as winners when the ELOC increase. 
Usually, larger size apps are more sophisticated and offer a richer set of user interactions, making their exploration more challenging for automated tools.
We already know from Study 1 that when the action space or the observation space of the apps increase, Deep RL can infer  the complex actions needed to explore such apps more easily than other algorithms. Study 2 confirms the same trend.

\begin{figure}[h!]
    \centering
    \includegraphics[width=0.8\textwidth]{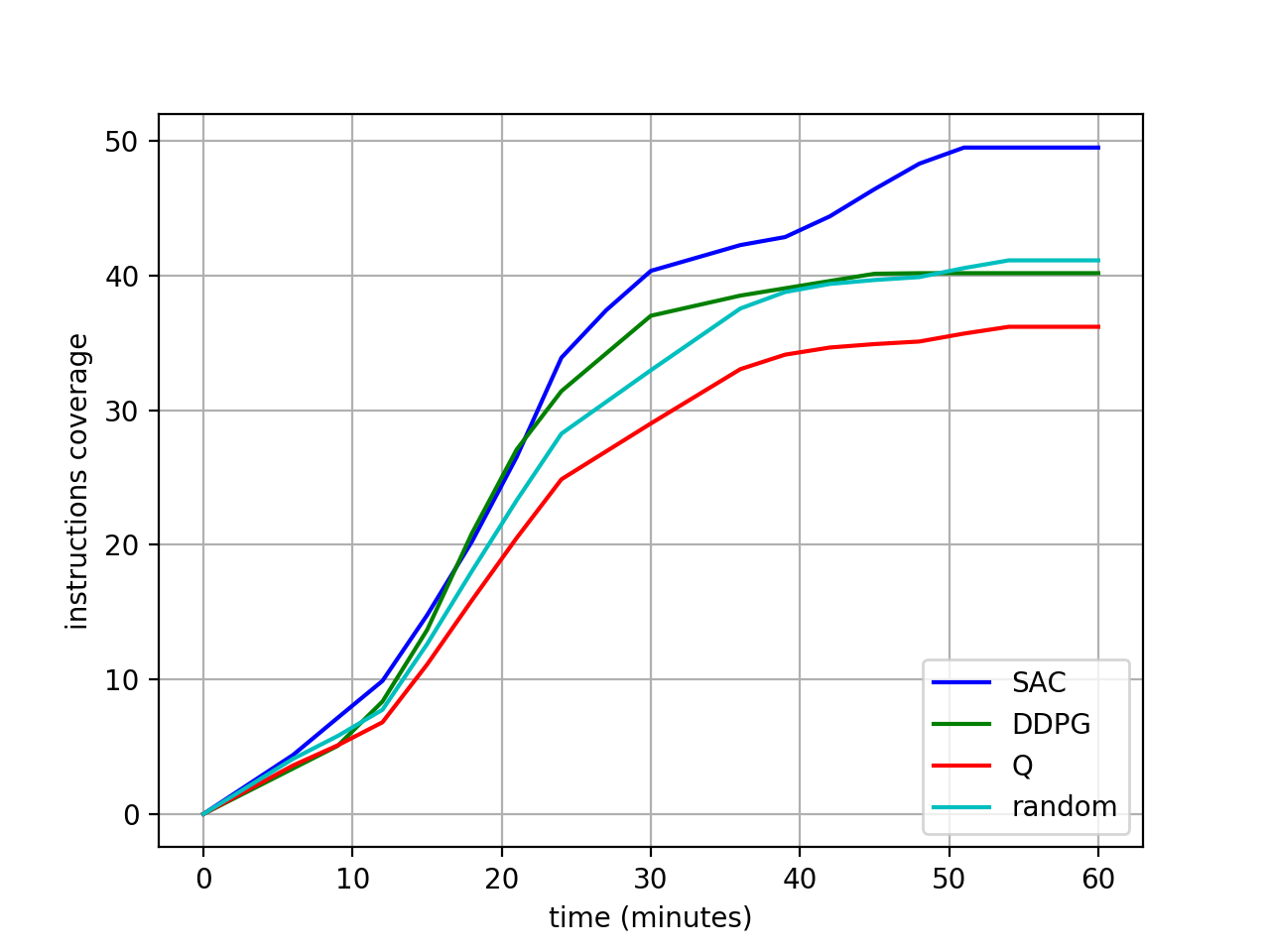}
    \caption{Instruction coverage over time for the app \textit{Loyalty-card-locker}}
    \label{fig:cov_time}
\end{figure}

Overall, SAC achieves the best performance, with 42.61\% instruction coverage and  0.3 faults detected on average. DDPG comes next, with 40.09\% instruction coverage and 0.12 faults detected on average.
To further investigate these results, we computed the AUCs reached by each algorithm and we applied the Wilcoxon test to each pair of algorithms. Table \ref{table:AUC} shows the AUCs achieved by the 4 algorithms and the Vargha-Delaney effect size between the winner and the other algorithms when the $p$-value is less than $\alpha$. SAC results as the winner on 56\% of the considered real apps, followed by Random (34\%). Moreover, Table \ref{table:AUC} confirms the trend observed in Table \ref{table:results}: as ELOC increase, a higher proportion of Deep RL algorithms produces the highest AUC.
Figure \ref{fig:cov_time} shows an example of  code coverage over time for the app \textit{Loyalty-card-locker}, averaged on 10 runs. SAC increases its coverage almost until the end of the exploration, while the other algorithms reach a plateau after around 35 minutes of exploration.

\begin{tcolorbox}[,colback=gray!5,colframe=blue!40, boxrule=0.3mm]
RQ5: SAC reached the highest coverage in 24/41 apps, followed by DDPG (11 apps), Random (10 apps), and Q-Learning(1 app). SAC has also the highest AUC in 24/41 apps, followed by Random (13 apps), Q-Learning (11 apps) and DDPG (5 apps).
\end{tcolorbox}

\begin{table}[h]
\centering
\scriptsize
\begin{tabularx}{0.61\textwidth}{|l|>{\centering\arraybackslash}m{0.45cm}>{\centering\arraybackslash}m{0.45cm}>{\centering\arraybackslash}m{0.55cm}>{\centering\arraybackslash}m{0.5cm}|>{\arraybackslash}m{2.4cm}|}
\hline
\textbf{App} & \textbf{Rand} & \textbf{Q} & \textbf{SAC} & \textbf{DDPG} & \textbf{Effect Size} \\
\hline
\hline
Silent-ping-sms & 0.36 & 0.36 & \cellcolor[HTML]{C0C0C0}0.38 & 0.37 & S(DDPG),M(Rand),L(Q)\\
Drawable-notepad & 0.13 & 0.14 & \cellcolor[HTML]{C0C0C0}0.20 & 0.18 & L(Rand,Q) \\
SmsMatrix & \cellcolor[HTML]{C0C0C0}0.13 & \cellcolor[HTML]{C0C0C0}0.13 & 0.11 & 0.11 & - \\
Busybox & 0.54 & 0.56 & \cellcolor[HTML]{C0C0C0}0.68 & \cellcolor[HTML]{C0C0C0}0.68 & L(Q,Rand) \\
WiFiKeyShare & 0.29 & 0.29 &\cellcolor[HTML]{C0C0C0} 0.33 & 0.30 & L(DDPG,Q,Rand) \\
Talalarmo & 0.62 & 0.60 & \cellcolor[HTML]{C0C0C0}0.64 &\cellcolor[HTML]{C0C0C0} 0.64 & L(Q) \\
AquaDroid & 0.529 & 0.526 & \cellcolor[HTML]{C0C0C0}0.531 & 0.522 & L(Rand, Q, DDPG) \\
Lexica & 0.63 & 0.61 & \cellcolor[HTML]{C0C0C0}0.66 & 0.65 & L(Q,Rand) \\
Loyalty-card-locker & 0.23 & 0.23 & \cellcolor[HTML]{C0C0C0}0.34 & 0.21 & L(DDPG,Q,Rand) \\
Dns66 & \cellcolor[HTML]{C0C0C0}0.51 & \cellcolor[HTML]{C0C0C0}0.51 & 0.47 & 0.45 & L(DDPG,SAC) \\
Gpstest & \cellcolor[HTML]{C0C0C0}0.40 & \cellcolor[HTML]{C0C0C0}0.40 & 0.39 & 0.36 & L(DDPG) \\
Memento & 0.64 & \cellcolor[HTML]{C0C0C0}0.65 & 0.64 & \cellcolor[HTML]{C0C0C0}0.65 & - \\
Editor & \cellcolor[HTML]{C0C0C0}0.42 & 0.37 & 0.37 & 0.37 & L(DDPG,Q,SAC)\\
AndOTP & 0.16 & 0.18 & \cellcolor[HTML]{C0C0C0}0.23 & 0.15 & M(Rand), L(DDPG) \\
BookyMcBookface & \cellcolor[HTML]{C0C0C0}0.22 & 0.20 & 0.20 & 0.20 & M(DDPG), L(SAC) \\
Tuner & \cellcolor[HTML]{C0C0C0}0.68 & 0.66 & 0.60 & 0.60 & L(DDPG,SAC) \\
WifiAnalyzer & 0.56 & 0.56 & \cellcolor[HTML]{C0C0C0}0.67 & 0.58 & L(DDPG,Q,Rand) \\
AdAway & 0.25 & 0.25 & \cellcolor[HTML]{C0C0C0}0.27 & 0.25 & -\\
Gpslogger & \cellcolor[HTML]{C0C0C0}0.28 & \cellcolor[HTML]{C0C0C0}0.28 & 0.23 & 0.20 & L(DDPG,SAC) \\
Connectbot & 0.19 & 0.19 & \cellcolor[HTML]{C0C0C0}0.22 & 0.09 & L(DDPG,Q,Rand) \\
Neurolab & \cellcolor[HTML]{C0C0C0}0.23 & \cellcolor[HTML]{C0C0C0} 0.23 & 0.22 & 0.22 & -\\
Anuto & 0.35 & 0.40 & \cellcolor[HTML]{C0C0C0}0.43 & 0.33 & L(DDPG,Q,Rand) \\
PassAndroid & \cellcolor[HTML]{C0C0C0}0.018 & 0.017 & 0.017 & 0.017 & - \\
Markor & 0.40 & 0.40 & \cellcolor[HTML]{C0C0C0}0.43 & 0.25 & L(DDPG,Q,Rand) \\
Vanilla & 0.17 & 0.23 & \cellcolor[HTML]{C0C0C0}0.26 & 0.23 & L(Rand) \\
\hline
\hline
Afwall & 0.09 & 0.09 & \cellcolor[HTML]{C0C0C0}0.13 & 0.10 & L(DDPG,Q,Rand) \\
OpenTracks & \cellcolor[HTML]{C0C0C0}0.37 & 0.35 & 0.35 & 0.35 & - \\
Opentasks & 0.18 & \cellcolor[HTML]{C0C0C0}0.46 & \cellcolor[HTML]{C0C0C0}0.46 & 0.29 & L(DDPG,Rand) \\
UserLAnd & \cellcolor[HTML]{C0C0C0}0.49 & \cellcolor[HTML]{C0C0C0}0.49 & \cellcolor[HTML]{C0C0C0}0.49 & 0.47 & -\\
Simple-Solitaire & 0.06 & \cellcolor[HTML]{C0C0C0}0.21 & 0.19 & 0.18 & L(Rand) \\
Authorizer & \cellcolor[HTML]{C0C0C0}0.05 & 0.046 & 0.047 & 0.049 & S(SAC), M(Q)\\
YalpStore & 0.28 & 0.28 & \cellcolor[HTML]{C0C0C0}0.31 & 0.26 & L(DDPG,Q,Rand) \\
Camera-Roll & 0.26 & \cellcolor[HTML]{C0C0C0}0.37 & 0.25 & 0.25 & L(Rand,DDPG,SAC) \\
AntennaPod & 0.33 & 0.33 & \cellcolor[HTML]{C0C0C0}0.35 & 0.22 & L(DDPG) \\
Phonograph & \cellcolor[HTML]{C0C0C0}0.085 & 0.077 & 0.075 & 0.076 & S(Q,DDPG,SAC) \\
\hline
\hline
MicroMathematics & 0.17 & 0.17 & \cellcolor[HTML]{C0C0C0}0.30 & 0.18 & L(DDPG,Q,Rand) \\
Lightning-Browser & 0.28 & 0.28 & \cellcolor[HTML]{C0C0C0}0.36 & 0.29 & L(DDPG,Q,Rand) \\
Firefox-focus & 0.27 & 0.28 & \cellcolor[HTML]{C0C0C0}0.43 & 0.35 & L(Rand, Q)\\
RedReader & 0.31 & 0.31 & 0.31 & \cellcolor[HTML]{C0C0C0}0.33 & -\\
Wikipedia & 0.30 & 0.30 & \cellcolor[HTML]{C0C0C0}0.33 & 0.28 & -\\
Slide & 0.13 & 0.13 & 0.12 & \cellcolor[HTML]{C0C0C0}0.14 & -\\
\hline
\end{tabularx}
\vspace{3mm}
\caption{AUCs achieved on real apps; effect size between winner and others when $p$-value $<$ $\alpha$}
\label{table:AUC}
\end{table}

\begin{figure}[h]
    \centering
    \includegraphics[width=0.8\textwidth]{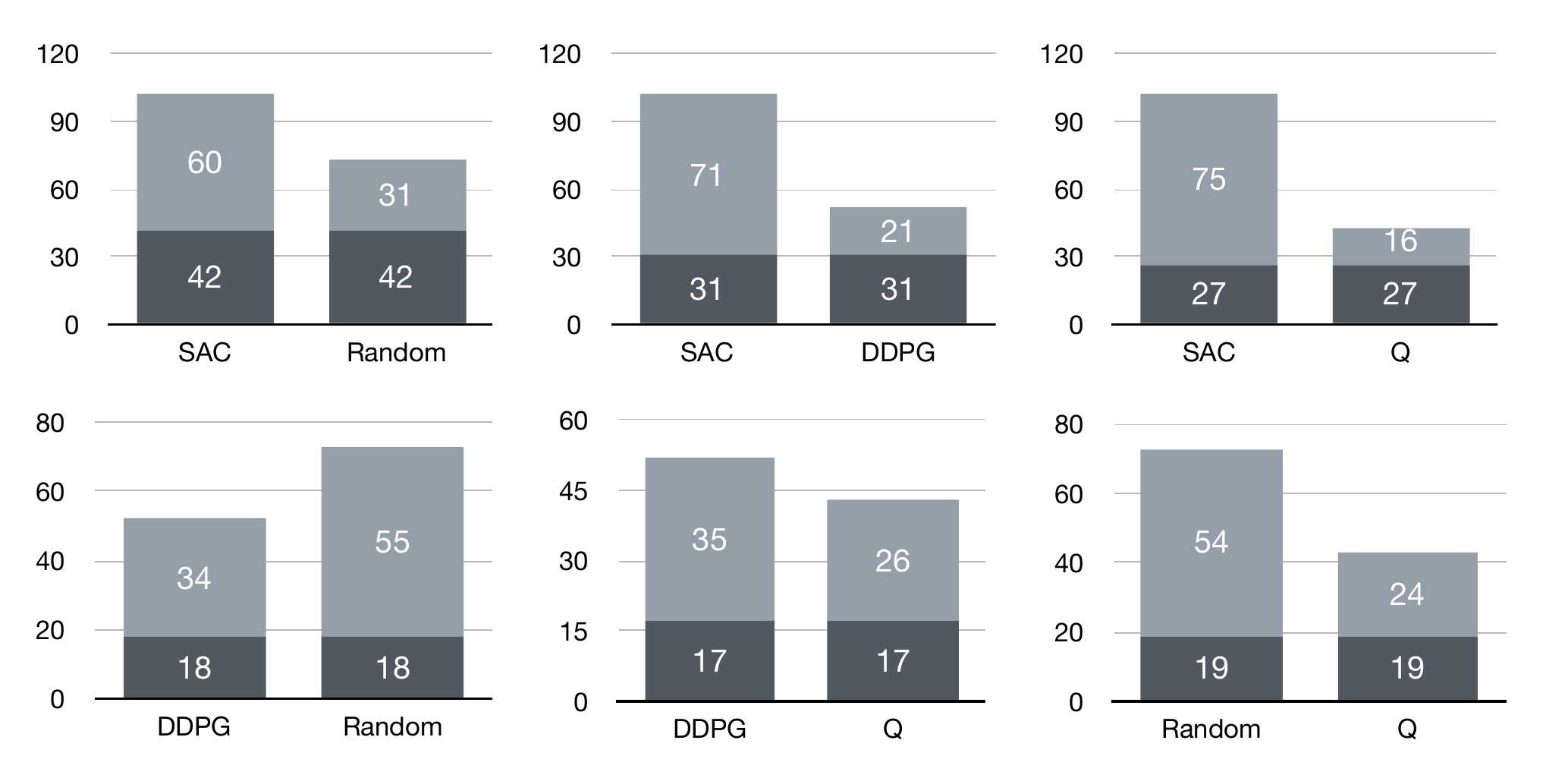}
    \caption{Comparison of total number of unique crashes on the 41 apps involved in RQ5-6-:  dark gray areas indicate the proportion of crashes found by both techniques}
    \label{fig:comparison}
\end{figure}

Table \ref{table:results} shows that SAC exposed the highest number of unique crashes (102), followed by Random (73), DDPG (52) and Q-Learning (43). The most common types of error exposed by SAC during testing are: \textit{RuntimeException} (34 occurrences), \textit{NullPointerException} (14), \textit{IllegalArgumentException} (13).
As shown in Figure \ref{fig:comparison}, there is around thirty percent overlap between the crashes found by SAC and the other algorithms. The overlap with Random is the highest. SAC discovers about 40\% of unique crashes found by Random; however, SAC found many new crashes that Random did not find.

%

\begin{tcolorbox}[,colback=gray!5,colframe=blue!40, boxrule=0.3mm]
RQ6: The SAC algorithm implemented in \toolname{} generates the highest number of crashes, 102, in line with the results on coverage (RQ5), where SAC was also the best performing algorithm.
\end{tcolorbox}

We have manually inspected the coverage progress of the different algorithms on some of the real apps considered in Study 2. We observed that deep RL algorithms achieve higher coverage than the other algorithms when it is necessary to replicate complex behaviors in order to: (1) overcome blocking activities, e.g., to create an item in order to  be able to later access its properties, or to bypass authentication; (2) to pass through concatenated activities without being distracted by already seen activities or ineffective buttons (high dimensional action/observation space); (3) reach an activity located deeply in the app.
Such behaviors are possible thanks to the learning capabilities of the DNNs used by deep RL algorithms, while they are hardly achieved by the other existing approaches, including tabular Q-Learning.

\begin{tcolorbox}[,colback=gray!5,colframe=blue!40, boxrule=0.3mm]
RQ7: In presence of blocking activities or complex concatenated activities (activities with a high number of widgets or located in depth) that require the capability to reuse knowledge acquired in previous explorations, the learning capabilities of deep RL algorithms make them the most effective and efficient exploration strategies.
\end{tcolorbox}
 

\subsubsection{Comparison between \toolname{} and state-of-the-art tools}

\begin{figure}[h]
    \centering
    \includegraphics[width=0.8\textwidth]{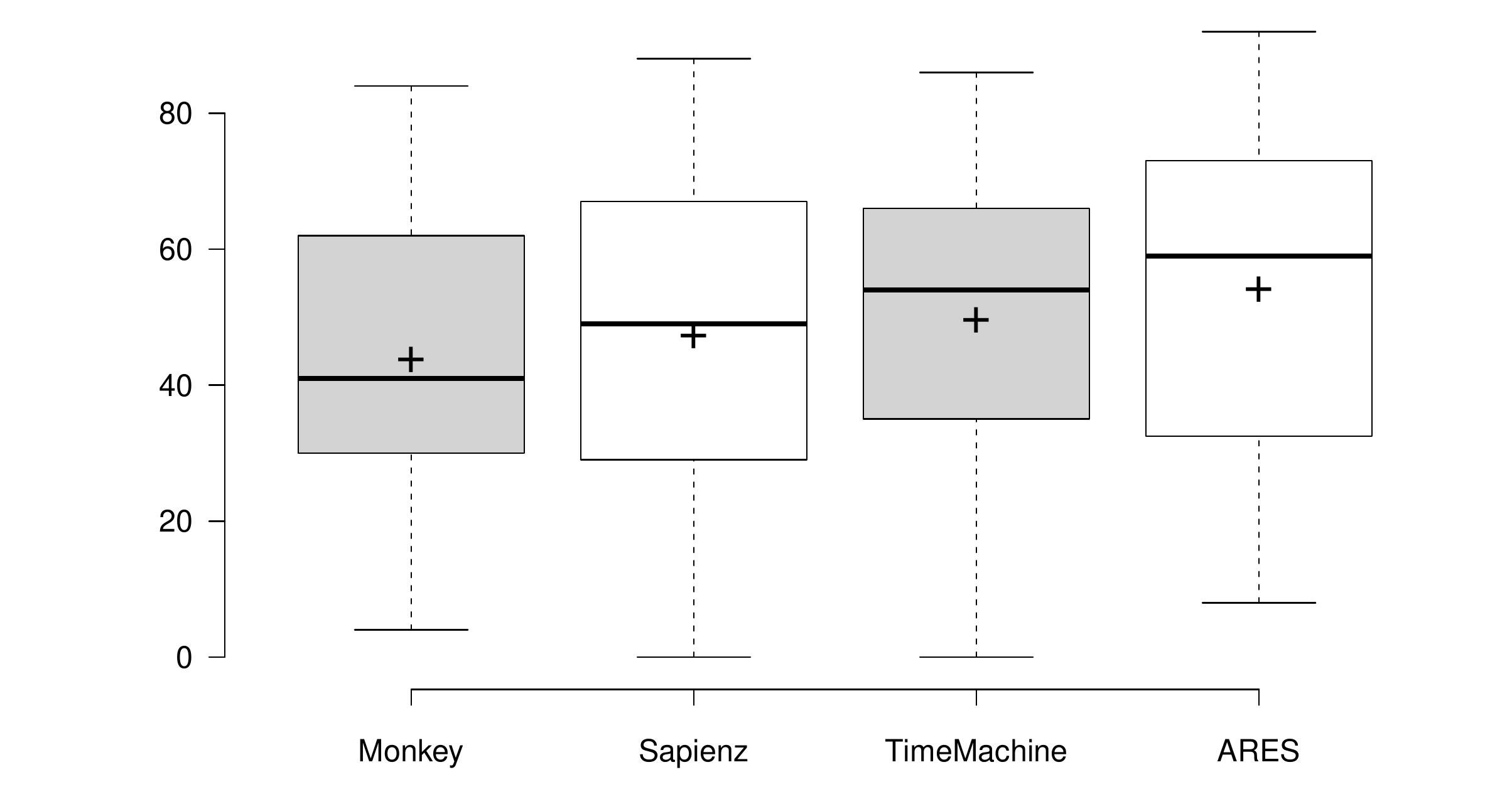}
    \caption{Code coverage achieved by \toolname{}, TimeMachine, Sapienz, and Monkey}
    \label{fig:comparison_sota2}
\end{figure}

\begin{figure}[h]
    \centering
    \includegraphics[width=0.8\textwidth]{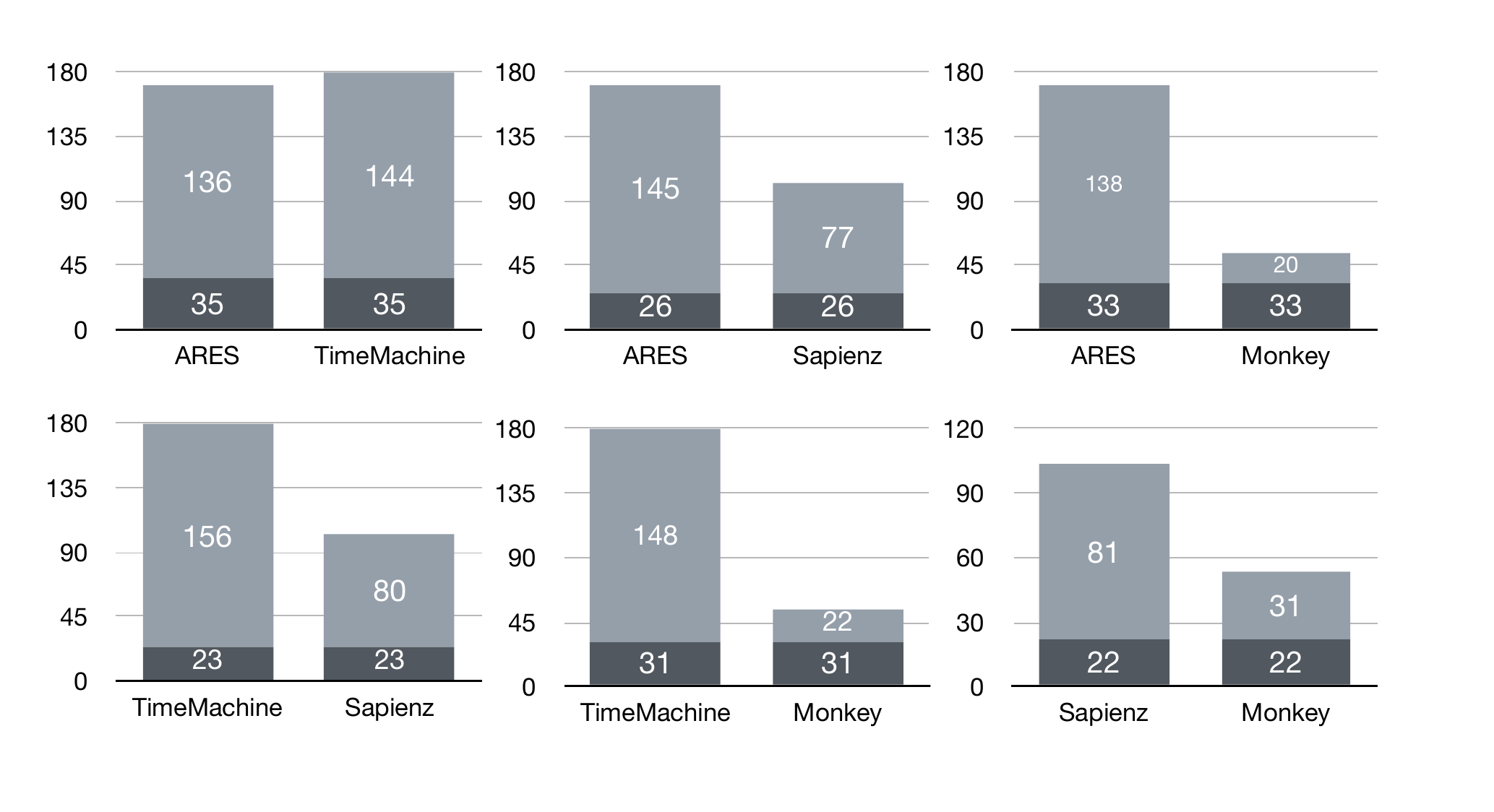}
    \caption{Comparison of total number of unique crashes involved in RQ8-a: dark gray areas indicate the proportion of crashes found by both testing tools}
    \label{fig:comparison_sota1}
\end{figure}

\textit{RQ8-a}. Table \ref{table:sota_comparison} shows the  coverage reached and the faults exposed by each testing tool on 68 Android apps from AndroTest.  Coverage data are summarized by means of boxplots in Figure~\ref{fig:comparison_sota2}. The highest average coverage and average number of crashes over 10 runs are highlighted with a gray background. \toolname{} achieved 54.2\% coverage and detected 0.48 crashes on average. TimeMachine achieved 50.4\% code coverage and 0.42 faults on average. Sapienz reached a mean code coverage of 48.8\%, and discovered 0.22 faults on average. Monkey achieved 43.9\% code coverage and discovered 0.11 faults.
\toolname{} achieved the highest code coverage on average,  followed by TimeMachine, Sapienz and Monkey.
TimeMachine detected  most unique crashes (179), followed by \toolname{} (171), Sapienz (103) and Monkey (51).
Actually, \toolname{} discovered less crashes than TimeMachine mostly because TimeMachine uses a system-level event generator, taken from Stoat \cite{DBLP:conf/sigsoft/SuMCWYYPLS17}, which \toolname{} does not support. However, system events present two major drawbacks: a) they vastly change depending on the Android version \cite{api19} \cite{api25} (despite TimeMachine is compatible with Android 7.1, it uses only system-level actions compatible with Android 4.4 \cite{timeMachinesys}); and b) in order to execute them, root privileges are required. ARES does not require root privileges in order to work properly on any app (i.e., we recall that protected apps do not execute on rooted devices~\cite{root_apps}).
Analyzing the execution traces of each tool, we searched and identified the faults immediately after the generation of system events not related to the AUT. More than a third (63) of the crashes generated by TimeMachine comes from system-level actions. Figure \ref{fig:comparison_sota1} shows a pairwise comparison of detected crashes among evaluated techniques. Only 20\% of unique crashes found by \toolname{} are also found by TimeMachine. This  shows that \toolname{} can be used together with other state-of-the-art Android testing techniques (in particular, TimeMachine) to jointly cover more code and discover more crashes.

The good results of \toolname{} in code coverage and exposed faults are due to the reinforcement mechanisms of the RL algorithms and to the reward function that drives  the testing tool through states of the app leading to hard-to-reach states.
Search-based techniques such as Sapienz typically observe the program behavior over an event sequence that may be very long. Hence, the associated coverage feedback, used to drive search-based exploration, does not have enough fine-grained details to support good local choices of the actions to be taken. In fact, the fitness function used by these algorithms evaluates an entire action sequence and does not consider the individual steps in the sequence. TimeMachine, improved this weakness, by relying on the coverage feedback obtained at an individual state to determine which portion of the app is not still explored.  The drawback of this kind of approach is a higher computational cost, that requires a higher testing time.
\toolname{} offers a better trade-off between the granularity of the feedback and the computational cost required to obtain it.

\begin{tcolorbox}[,colback=gray!5,colframe=blue!40, boxrule=0.3mm]
RQ8-a: \toolname{} achieved the highest code coverage in 41/68 apps, followed by TimeMachine (12/68), Sapienz (6/68) and Monkey (2/68). \toolname{} triggered crashes more often than other tools in 23/68 apps, followed by TimeMachine (15/68), Sapienz (5/68) and Monkey (0/68). However, TimeMachine generated the highest number of unique crashes (179), but 63 of them come from system-level events. \toolname{} generates 171 faults, Sapienz 103, and Monkey 51.
\end{tcolorbox}

\begin{table}[]
\centering
\tiny
\begin{tabular}{|c|cccc|cccc|}
\hline
\textbf{App} & \multicolumn{4}{c|}{\textbf{Coverage}} & \multicolumn{4}{c|}{\textbf{Faults}} \\
\textbf{} & \textbf{Monkey} & \textbf{Sapienz} & \textbf{TM} & \textbf{ARES} & \textbf{Monkey} & \textbf{Sapienz} & \textbf{TM} & \textbf{ARES} \\
\hline
a2dp & 38 & \cellcolor[HTML]{C0C0C0}44 & 42 & 42 & 0 & \cellcolor[HTML]{C0C0C0}0.4 & 0.1 & 0.3 \\
aagtl & 16 & 18 & 17 & \cellcolor[HTML]{C0C0C0}19 & 0.3 & 1.0 & \cellcolor[HTML]{C0C0C0}1.7 & 1.5 \\
aarddict & 13 & 15 & 17 & 17 & 0 & 0 & \cellcolor[HTML]{C0C0C0}0.2 & \cellcolor[HTML]{C0C0C0}0.2 \\
acal & 18 & 27 & 27 & \cellcolor[HTML]{C0C0C0}28 & 0.5 & 0.8 & \cellcolor[HTML]{C0C0C0}1.0 & \cellcolor[HTML]{C0C0C0}1.0 \\
addi & 19 & 20 & 17 & \cellcolor[HTML]{C0C0C0}20 & 0.4 & 0.3 & 0.4 & \cellcolor[HTML]{C0C0C0}0.6 \\
adsdroid & 30 & \cellcolor[HTML]{C0C0C0}36 & \cellcolor[HTML]{C0C0C0}36 & 35 & 0.1 & 0.4 & 0.5 & \cellcolor[HTML]{C0C0C0}0.7 \\
aGrep & 45 & - & \cellcolor[HTML]{C0C0C0}59 & 56 & 0.1 & - & \cellcolor[HTML]{C0C0C0}0.3 & 0.2 \\
aka & 65 & 84 & 77 & \cellcolor[HTML]{C0C0C0}84 & 0.1 & \cellcolor[HTML]{C0C0C0}0.7 & 0.1 & 0.4 \\
alarmclock & 64 & 41 & 60 & \cellcolor[HTML]{C0C0C0}71 & 0.5 & 0.5 & 0.6 & \cellcolor[HTML]{C0C0C0}0.8 \\
aLogCat & 67 & 71 & 75 & \cellcolor[HTML]{C0C0C0}87 & 0 & 0 & 0 & 0 \\
Amazed & 36 & 66 & 63 & \cellcolor[HTML]{C0C0C0}89 & 0.1 & \cellcolor[HTML]{C0C0C0}0.3 & 0.2 & 0.2 \\
AnyCut & 63 & 65 & 63 & \cellcolor[HTML]{C0C0C0}73 & 0 & 0 & 0 & 0 \\
anymemo & 31 & 50 & 43 & \cellcolor[HTML]{C0C0C0}53 & 0.2 & 0.8 & 0.3 & \cellcolor[HTML]{C0C0C0}1.0 \\
autoanswer & 12 & \cellcolor[HTML]{FFFFFF}16 & \cellcolor[HTML]{C0C0C0}21 & 15 & 0 & 0 & \cellcolor[HTML]{C0C0C0}0.5 & \cellcolor[HTML]{C0C0C0}0.5 \\
baterrydog & 63 & 67 & 62 & \cellcolor[HTML]{C0C0C0}69 & 0 & 0.1 & 0.4 & \cellcolor[HTML]{C0C0C0}0.5 \\
battery & 73 & 78 & 77 & \cellcolor[HTML]{C0C0C0}92 & 0 & 0.4 & \cellcolor[HTML]{C0C0C0}0.5 & 0.4 \\
bites & 34 & 41 & \cellcolor[HTML]{C0C0C0}45 & 43 & 0.1 & 0.2 & \cellcolor[HTML]{C0C0C0}0.9 & 0.5 \\
blokish & 55 & 52 & \cellcolor[HTML]{C0C0C0}68 & 45 & 0 & 0.2 & 0 & \cellcolor[HTML]{C0C0C0}0.3 \\
bomber & 76 & 73 & 77 & \cellcolor[HTML]{C0C0C0}84 & 0 & 0 & 0 & 0 \\
Book-Catalogue & 27 & \cellcolor[HTML]{C0C0C0}29 & 27 & 25 & 0.1 & 0.2 & \cellcolor[HTML]{C0C0C0}0.8 & 0.5 \\
CountdownTimer & 74 & 62 & 77 & \cellcolor[HTML]{C0C0C0}84 & 0 & 0 & 0 & 0 \\
dalvik-explorer & 66 & \cellcolor[HTML]{C0C0C0}72 & 70 & \cellcolor[HTML]{C0C0C0}72 & 0.3 & 0.2 & 0.7 & \cellcolor[HTML]{C0C0C0}0.8 \\
dialer2 & 39 & 42 & 42 & \cellcolor[HTML]{C0C0C0}44 & 0 & 0 & 0.3 & \cellcolor[HTML]{C0C0C0}0.4 \\
DivideAndConquer & \cellcolor[HTML]{C0C0C0}84 & 83 & 82 & 80 & 0 & 0.2 & \cellcolor[HTML]{C0C0C0}1.0 & 0.9 \\
fileexplorer & 41 & 49 & 55 & \cellcolor[HTML]{C0C0C0}64 & 0 & 0 & 0 & 0 \\
frozenbubble & \cellcolor[HTML]{C0C0C0}80 & 76 & 75 & 70 & 0 & 0 & 0 & 0 \\
gestures & 37 & 52 & 51 & \cellcolor[HTML]{C0C0C0}55 & 0 & 0 & 0 & 0 \\
hndroid & 7 & 15 & \cellcolor[HTML]{C0C0C0}18 & \cellcolor[HTML]{C0C0C0}18 & 0.1 & 0.4 & 1.1 & \cellcolor[HTML]{C0C0C0}1.3 \\
hotdeath & \cellcolor[HTML]{C0C0C0}75 & \cellcolor[HTML]{C0C0C0}75 & 72 & 74 & 0.1 & 0.2 & 0.8 & \cellcolor[HTML]{C0C0C0}0.9 \\
importcontacts & 40 & 39 & 40 & \cellcolor[HTML]{C0C0C0}42 & 0.1 & 0 & 0.6 & \cellcolor[HTML]{C0C0C0}0.8 \\
jamendo & 53 & 41 & 54 & \cellcolor[HTML]{C0C0C0}63 & 0 & 0.4 & 1.4 & \cellcolor[HTML]{C0C0C0}1.6 \\
k9mail & 6 & 7 & \cellcolor[HTML]{C0C0C0}8 & \cellcolor[HTML]{C0C0C0}8 & 0.4 & 0 & \cellcolor[HTML]{C0C0C0}1.8 & 1.2 \\
LNM & 47 & - & - & \cellcolor[HTML]{C0C0C0}75 & 0 & - & - & \cellcolor[HTML]{C0C0C0}0.2 \\
lockpatterngenerator & 75 & \cellcolor[HTML]{C0C0C0}79 & 74 & 78 & 0 & 0 & 0 & 0 \\
LolcatBuilder & 26 & 25 & \cellcolor[HTML]{C0C0C0}29 & 26 & 0 & 0 & \cellcolor[HTML]{C0C0C0}0.1 & 0 \\
manpages & 40 & 73 & 70 & \cellcolor[HTML]{C0C0C0}74 & 0 & \cellcolor[HTML]{C0C0C0}0.4 & 0.3 & \cellcolor[HTML]{C0C0C0}0.4 \\
mileage & 38 & 45 & \cellcolor[HTML]{C0C0C0}48 & 45 & 0.3 & 1.0 & \cellcolor[HTML]{C0C0C0}2.3 & 1.8 \\
Mirrored & 57 & 59 & \cellcolor[HTML]{C0C0C0}62 & 59 & 0.4 & 0 & \cellcolor[HTML]{C0C0C0}0.8 & 0.7 \\
mnv & 41 & \cellcolor[HTML]{C0C0C0}60 & 43 & 56 & 0.5 & 0.3 & 1.0 & \cellcolor[HTML]{C0C0C0}1.1 \\
multismssender & 34 & 59 & 61 & \cellcolor[HTML]{C0C0C0}73 & 0 & 0.2 & 0.3 & \cellcolor[HTML]{C0C0C0}0.4 \\
MunchLife & 67 & 72 & 71 & \cellcolor[HTML]{C0C0C0}88 & 0 & 0 & 0 & 0 \\
MyExpenses & 41 & 60 & 50 & \cellcolor[HTML]{C0C0C0}63 & 0 & 0 & \cellcolor[HTML]{C0C0C0}0.2 & \cellcolor[HTML]{C0C0C0}0.2 \\
myLock & 25 & 31 & \cellcolor[HTML]{C0C0C0}50 & 30 & 0 & 0 & \cellcolor[HTML]{C0C0C0}0.5 & 0.2 \\
Nectroid & 34 & \cellcolor[HTML]{C0C0C0}66 & 58 & 57 & 0 & \cellcolor[HTML]{C0C0C0}1.0 & 0.3 & 0.9 \\
netcounter & 43 & \cellcolor[HTML]{C0C0C0}70 & 58 & 69 & 0 & 0 & 0.4 & \cellcolor[HTML]{C0C0C0}0.5 \\
PasswordMaker & 53 & 58 & 55 & \cellcolor[HTML]{C0C0C0}59 & 0.3 & 0.8 & 0.6 & \cellcolor[HTML]{C0C0C0}0.9 \\
passwordmanager & 7 & 8 & 17 & \cellcolor[HTML]{C0C0C0}18 & 0 & 0 & 0 & 0 \\
Photostream & 30 & 34 & \cellcolor[HTML]{C0C0C0}35 & 29 & 0.1 & 0 & 0.4 & \cellcolor[HTML]{C0C0C0}0.8 \\
QuickSettings & 50 & 45 & 46 & \cellcolor[HTML]{C0C0C0}52 & 0 & 0.2 & 0 & \cellcolor[HTML]{C0C0C0}0.4 \\
RandomMusicPlayer & 53 & 58 & 58 & \cellcolor[HTML]{C0C0C0}63 & 0 & 0 & 0.6 & \cellcolor[HTML]{C0C0C0}0.8 \\
Ringdroid & 22 & 29 & \cellcolor[HTML]{C0C0C0}48 & 30 & 0 & 0.1 & \cellcolor[HTML]{C0C0C0}0.3 & 0.2 \\
sanity & 26 & 19 & \cellcolor[HTML]{C0C0C0}31 & 22 & 0.2 & 0.3 & \cellcolor[HTML]{C0C0C0}0.5 & 0.4 \\
soundboard & 42 & 32 & 59 & \cellcolor[HTML]{C0C0C0}61 & 0 & \cellcolor[HTML]{C0C0C0}0.6 & 0 & 0.4 \\
SpriteMethodTest & 58 & 80 & 73 & \cellcolor[HTML]{C0C0C0}88 & 0 & 0 & 0 & 0 \\
SpriteText & \cellcolor[HTML]{C0C0C0}60 & \cellcolor[HTML]{C0C0C0}60 & 57 & \cellcolor[HTML]{C0C0C0}60 & 0 & 0.4 & 0 & \cellcolor[HTML]{C0C0C0}0.5 \\
swiftp & 12 & 14 & 13 & \cellcolor[HTML]{C0C0C0}17 & 0 & 0.4 & - & \cellcolor[HTML]{C0C0C0}0.6 \\
SyncMyPix & 21 & 21 & 23 & \cellcolor[HTML]{C0C0C0}25 & 0 & 0 & 0 & 0 \\
tippy & 75 & 83 & 74 & \cellcolor[HTML]{C0C0C0}85 & 0 & \cellcolor[HTML]{C0C0C0}0.4 & 0.3 & \cellcolor[HTML]{C0C0C0}0.4 \\
tomdroid & 47 & 46 & 51 & \cellcolor[HTML]{C0C0C0}69 & 0 & \cellcolor[HTML]{C0C0C0}0.3 & 0 & \cellcolor[HTML]{C0C0C0}0.3 \\
Translate & 49 & 48 & 48 & \cellcolor[HTML]{C0C0C0}50 & 0 & 0 & 0 & 0 \\
Triangle & - & - & - & - & - & - & - & - \\
weight-chart & 63 & 67 & 66 & \cellcolor[HTML]{C0C0C0}71 & 0 & 0 & 0 & 0 \\
whohasmystuff & 61 & 68 & 66 & \cellcolor[HTML]{C0C0C0}81 & 0.1 & 0 & 0.9 & \cellcolor[HTML]{C0C0C0}1.0 \\
wikipedia & 31 & 32 & 33 & \cellcolor[HTML]{C0C0C0}35 & 0 & 0 & 0 & 0 \\
Wordpress & 4 & 5 & 7 & \cellcolor[HTML]{C0C0C0}8 & 0 & 0.5 & \cellcolor[HTML]{C0C0C0}1.5 & 1.0 \\
worldclock & 83 & 88 & 86 & \cellcolor[HTML]{C0C0C0}90 & 0 & 0 & \cellcolor[HTML]{C0C0C0}0.6 & 0.4 \\
yahtzee & 51 & 57 & 56 & \cellcolor[HTML]{C0C0C0}69 & 2 & 0.2 & \cellcolor[HTML]{C0C0C0}0.5 & \cellcolor[HTML]{C0C0C0}0.5 \\
zooborns & 30 & 16 & 35 & \cellcolor[HTML]{C0C0C0}37 & 0 & 0 & 0.1 & \cellcolor[HTML]{C0C0C0}0.5 \\
\hline
\textbf{Average} & 43.9 & 48.8 & 50.4 & 54.2 & 0.11 & 0.22 & 0.42 & 0.48 \\
\textbf{Sum} & & & & & 51 & 103 & 179 & 171\\
\hline
\end{tabular}
\vspace{3mm}
\caption{Results on 68 open-source apps coming from AndroTest}
\label{table:sota_comparison}
\end{table}

\newpage

\textit{RQ8-b}. Table \ref{table:q_test_comparison} shows  the coverage reached and the faults exposed by \toolname{} and Q-Testing on 10 Android apps instrumented with JaCoCo. \toolname{} achieved 64.3\% coverage and exposed 0.41 faults on average; it detected 17 unique crashes. Q-Testing achieved 52.5\% code coverage and 0.27 crashes on average, and it detected 16 unique crashes.
\toolname{} achieved the highest code coverage on almost all apps, and on average \toolname{} covered 12\% more code than Q-Testing. Q-testing generated 6 faults in common with \toolname{}, while other 4 faults are generated using the system-level events of Stoat.
In this comparison, the main advantage of \toolname{} seems to be a better reward function, that encourages the tool to visit the greatest number of activities within the same episode. Instead, Q-Testing determines the reward of the Q-Learning algorithm by computing the similarity between Android app states, which does not guarantee an efficient way to overcome blocking activities.

\begin{figure}[h]
    \centering
    \includegraphics[width=0.8\textwidth]{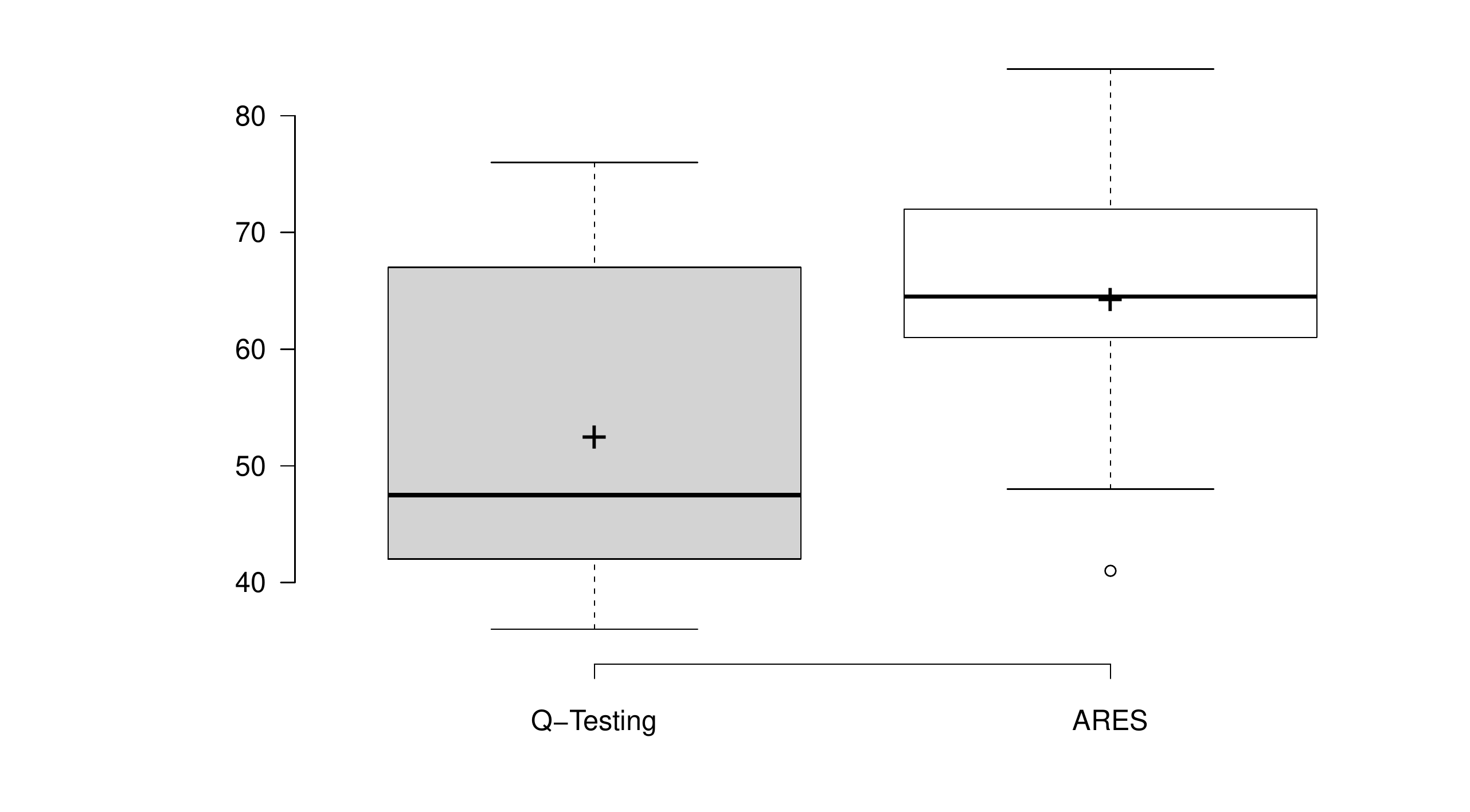}
    \caption{Code coverage achieved by \toolname{} and Q-Testing.}
    \label{fig:comparison_q_test}
\end{figure}

\begin{table}[]
\centering
\scriptsize
\begin{tabular}{|c|cc|cc|}
\hline
\textbf{App} & \multicolumn{2}{c|}{\textbf{Coverage}} & \multicolumn{2}{c|}{\textbf{Faults}} \\
 & \textbf{ARES} & \textbf{Q-Testing} & \textbf{ARES} & \textbf{Q-Testing} \\
 \hline
Alogcat & \cellcolor[HTML]{C0C0C0}84 & 76 & 0 & 0 \\
Antennapod & \cellcolor[HTML]{C0C0C0}48 & 42 & \cellcolor[HTML]{C0C0C0}0.8 & 0.4 \\
AnyCut & \cellcolor[HTML]{C0C0C0}72 & 67 & 0 & 0 \\
batterydog & \cellcolor[HTML]{C0C0C0}65 & 49 & \cellcolor[HTML]{C0C0C0}0.5 & 0.3 \\
Jamendo & \cellcolor[HTML]{C0C0C0}64 & 46 & 0.4 & \cellcolor[HTML]{C0C0C0}0.7 \\
Multismssender & \cellcolor[HTML]{C0C0C0}71 & 45 & \cellcolor[HTML]{C0C0C0}0.4 & 0 \\
Myexpanses & \cellcolor[HTML]{C0C0C0}63 & 36 & \cellcolor[HTML]{C0C0C0}0.4 & 0.2 \\
talalarmo & 74 & 74 & \cellcolor[HTML]{C0C0C0}0.8 & 0.5 \\
Tomdroid & \cellcolor[HTML]{C0C0C0}61 & 50 & \cellcolor[HTML]{C0C0C0}0.3 & 0.2 \\
vanilla & \cellcolor[HTML]{C0C0C0}41 & 40 & \cellcolor[HTML]{C0C0C0}0.5 & 0.4 \\
\hline
\textbf{Average} & 64.3 & 52.5 & 0.41 & 0.27 \\
\textbf{Sum} &  &  & 17 & 16 \\
\hline
\end{tabular}
\vspace{3mm}
\caption{Comparison between Q-Testing and ARES}
\label{table:q_test_comparison}
\end{table}

\begin{tcolorbox}[,colback=gray!5,colframe=blue!40, boxrule=0.3mm]
RQ8-b: \toolname{} reached the highest code coverage on 9/10 apps, the highest average of crashes in 7/10 apps, and the highest number of unique crashes. Q-Testing, the state-of-the art RL Android testing tool, reached the same level of exploration of \toolname{} only in 1/10 app, and the highest average number of crashes on only 1/10 apps.
\end{tcolorbox}
\newpage
\section{Related Work}


\subsection{Random Testing} Testing tools in this category generate random events on the AUT GUI.
Monkey \cite{Monkey} is one of the most popular black-box Android testing tools. It triggers events by interacting randomly with screen coordinates. This simple random approach worked relatively well on some benchmark applications \cite{automated}.
Nonetheless, Monkey tests involve a significant number of ineffective or repeated events, as there is no guidance to make the exploration efficient. \toolname{} implements a smarter version of Monkey, which we used as a baseline (Random). Such improved version selects  only  actions that are actually possible in a given GUI state, thus making the exploration strategy a bit more efficient.

\subsection{Model-based Testing} Model-based tools \cite{amalfitano2012using} \cite{amalfitano2014mobiguitar} first build navigation models of the Android app by means of static or dynamic analysis, used to  explore efficiently the application states, and then they extract test cases from such models, to eventually expose bugs. Stoat \cite{DBLP:conf/sigsoft/SuMCWYYPLS17} uses a stochastic FSM to model the app behavior. The app model is built using dynamic analysis, enhanced with a weighted UI exploration strategy and with the help of static analysis. Compared to Stoat, \toolname{} can be viewed as computing a navigation model implicitly: the MDP model used by the deep RL algorithms. One key advantage of using an implicit model is that we do not have to deal with the combinatorial explosion of its size, which is controlled by Stoat by means of model compaction heuristics.

\subsection{Structural Testing} Structural strategies \cite{anand2012automated,gao2018android,mahmood2014evodroid,dong2020timetravel,mao2016sapienz} generate coverage oriented inputs using symbolic execution or evolutionary algorithms. Sapienz \cite{mao2016sapienz} maximizes code coverage and bug revelation using a Pareto-optimal multi-objective search based approach, which applies genetic operators such as mutation and crossover to produce new test cases. It can  generate specific input for text fields by reverse engineering the APK. This process occasionally results in invalid sequences, which are discarded by the fitness functions that reward test cases with high coverage. 
TimeMachine~\cite{dong2020timetravel} improves Sapienz by identifying interesting states in the past and restarting the search process from them when the search stagnates.
While coverage oriented approaches require the possibility to instrument the app to measure coverage and to possibly execute it symbolically, our approach is black-box and does not suffer from the scalability issues affecting, e.g., symbolic execution.

\subsection{Machine Learning Based Testing}
Some ML based testing approaches \cite{borges2018guiding} \cite{koroglu2018qbe} \cite{li2019humanoid} use an explicit, supervised training process to learn from previous test executions. They can reuse previous knowledge, acquired on different apps or on past versions of the app under test.
Approaches as QBE \cite{koroglu2018qbe} make  the transfer of knowledge to new apps possible by abstracting the app state in a form that is supposed to hold across different domains and implementations. However, the effectiveness of such transfer learning process depends on the similarity between new and old apps.

One of the first works proposing RL for GUI testing is AutoBlackTest \cite{mariani2012autoblacktest}. This approach is based on the simplest form of RL, tabular Q-Learning, whose effectiveness is strongly dependent on the initial values in the Q-Table. On the contrary, \toolname{} learns the action-value function from scratch during the exploration of the AUT. One of the most recent approaches to Android testing based on deep learning is Q-Testing \cite{qtest}. However, it also uses tabular Q-Learning as backbone, while  learning is limited to the computation of the similarity between Android app states, which determines the reward of the Q-Learning algorithm. \toolname{} instead learns both the state similarity and the action-value function during its interactions with the AUT.

\section{Conclusion and Future Work} \label{sec:concl}

We have proposed an approach based on Deep RL for the automated exploration of Android apps. The best exploration strategy is learned automatically as the test progresses. The approach is implemented in the open source tool \toolname{}, which is complemented by FATE, a model based Android testing tool that we developed to support fast execution and configuration of the alternative deep RL algorithms of \toolname{}. 
The resulting configuration of \toolname{}, in particular when running the SAC deep RL algorithm, outperformed all the considered baselines  in terms of  coverage achieved over time and  exposed bugs.

In our future work, we plan to investigate specific fault categories that are particularly relevant for Android apps, such as security vulnerabilities. In fact, we think that the adaptation and reward mechanisms used by deep RL algorithms to learn the optimal exploration strategy could be particularly effective when the fault to be exposed is a security fault.
We also plan to port \toolname{} and FATE to iOS.

\bibliographystyle{ACM-Reference-Format}
\bibliography{bibliography}


\begin{thebibliography}{43}


\ifx \showCODEN    \undefined \def \showCODEN     #1{\unskip}     \fi
\ifx \showDOI      \undefined \def \showDOI       #1{#1}\fi
\ifx \showISBNx    \undefined \def \showISBNx     #1{\unskip}     \fi
\ifx \showISBNxiii \undefined \def \showISBNxiii  #1{\unskip}     \fi
\ifx \showISSN     \undefined \def \showISSN      #1{\unskip}     \fi
\ifx \showLCCN     \undefined \def \showLCCN      #1{\unskip}     \fi
\ifx \shownote     \undefined \def \shownote      #1{#1}          \fi
\ifx \showarticletitle \undefined \def \showarticletitle #1{#1}   \fi
\ifx \showURL      \undefined \def \showURL       {\relax}        \fi
\providecommand\bibfield[2]{#2}
\providecommand\bibinfo[2]{#2}
\providecommand\natexlab[1]{#1}
\providecommand\showeprint[2][]{arXiv:#2}

\bibitem[\protect\citeauthoryear{??}{emm}{2006}]%
        {emma}
 \bibinfo{year}{2006}\natexlab{}.
\newblock \bibinfo{title}{Emma}.
\newblock
\newblock
\urldef\tempurl%
\url{http://emma.sourceforge.net}
\showURL{%
Retrieved December 30, 2020 from \tempurl}


\bibitem[\protect\citeauthoryear{??}{app}{2020}]%
        {appbrain}
 \bibinfo{year}{2020}\natexlab{}.
\newblock \bibinfo{title}{Appbrain}.
\newblock
\newblock
\urldef\tempurl%
\url{https://www.appbrain.com}
\showURL{%
Retrieved November 20, 2020 from \tempurl}


\bibitem[\protect\citeauthoryear{??}{App}{2020}]%
        {Appium}
 \bibinfo{year}{2020}\natexlab{}.
\newblock \bibinfo{title}{Appium}.
\newblock
\newblock
\urldef\tempurl%
\url{http://appium.io}
\showURL{%
Retrieved September 25, 2020 from \tempurl}


\bibitem[\protect\citeauthoryear{Amalfitano, Fasolino, Tramontana, De~Carmine,
  and Memon}{Amalfitano et~al\mbox{.}}{2012}]%
        {amalfitano2012using}
\bibfield{author}{\bibinfo{person}{Domenico Amalfitano},
  \bibinfo{person}{Anna~Rita Fasolino}, \bibinfo{person}{Porfirio Tramontana},
  \bibinfo{person}{Salvatore De~Carmine}, {and} \bibinfo{person}{Atif~M
  Memon}.} \bibinfo{year}{2012}\natexlab{}.
\newblock \showarticletitle{Using GUI ripping for automated testing of Android
  applications}. In \bibinfo{booktitle}{\emph{2012 Proceedings of the 27th
  IEEE/ACM International Conference on Automated Software Engineering}}. IEEE,
  \bibinfo{pages}{258--261}.
\newblock


\bibitem[\protect\citeauthoryear{Amalfitano, Fasolino, Tramontana, Ta, and
  Memon}{Amalfitano et~al\mbox{.}}{2014}]%
        {amalfitano2014mobiguitar}
\bibfield{author}{\bibinfo{person}{Domenico Amalfitano},
  \bibinfo{person}{Anna~Rita Fasolino}, \bibinfo{person}{Porfirio Tramontana},
  \bibinfo{person}{Bryan~Dzung Ta}, {and} \bibinfo{person}{Atif~M Memon}.}
  \bibinfo{year}{2014}\natexlab{}.
\newblock \showarticletitle{MobiGUITAR: Automated model-based testing of mobile
  apps}.
\newblock \bibinfo{journal}{\emph{IEEE software}} \bibinfo{volume}{32},
  \bibinfo{number}{5} (\bibinfo{year}{2014}), \bibinfo{pages}{53--59}.
\newblock


\bibitem[\protect\citeauthoryear{Anand, Naik, Harrold, and Yang}{Anand
  et~al\mbox{.}}{2012}]%
        {anand2012automated}
\bibfield{author}{\bibinfo{person}{Saswat Anand}, \bibinfo{person}{Mayur Naik},
  \bibinfo{person}{Mary~Jean Harrold}, {and} \bibinfo{person}{Hongseok Yang}.}
  \bibinfo{year}{2012}\natexlab{}.
\newblock \showarticletitle{Automated concolic testing of smartphone apps}. In
  \bibinfo{booktitle}{\emph{Proceedings of the ACM SIGSOFT 20th International
  Symposium on the Foundations of Software Engineering}}.
  \bibinfo{pages}{1--11}.
\newblock


\bibitem[\protect\citeauthoryear{Borges, G{\'o}mez, and Zeller}{Borges
  et~al\mbox{.}}{2018}]%
        {borges2018guiding}
\bibfield{author}{\bibinfo{person}{Nataniel~P Borges}, \bibinfo{person}{Maria
  G{\'o}mez}, {and} \bibinfo{person}{Andreas Zeller}.}
  \bibinfo{year}{2018}\natexlab{}.
\newblock \showarticletitle{Guiding app testing with mined interaction models}.
  In \bibinfo{booktitle}{\emph{2018 IEEE/ACM 5th International Conference on
  Mobile Software Engineering and Systems (MOBILESoft)}}. IEEE,
  \bibinfo{pages}{133--143}.
\newblock


\bibitem[\protect\citeauthoryear{Boyan and Moore}{Boyan and Moore}{1995}]%
        {boyan1995generalization}
\bibfield{author}{\bibinfo{person}{Justin~A Boyan} {and}
  \bibinfo{person}{Andrew~W Moore}.} \bibinfo{year}{1995}\natexlab{}.
\newblock \showarticletitle{Generalization in reinforcement learning: Safely
  approximating the value function}. In \bibinfo{booktitle}{\emph{Advances in
  neural information processing systems}}. \bibinfo{pages}{369--376}.
\newblock


\bibitem[\protect\citeauthoryear{Brockman, Cheung, Pettersson, Schneider,
  Schulman, Tang, and Zaremba}{Brockman et~al\mbox{.}}{2016}]%
        {gym}
\bibfield{author}{\bibinfo{person}{Greg Brockman}, \bibinfo{person}{Vicki
  Cheung}, \bibinfo{person}{Ludwig Pettersson}, \bibinfo{person}{Jonas
  Schneider}, \bibinfo{person}{John Schulman}, \bibinfo{person}{Jie Tang},
  {and} \bibinfo{person}{Wojciech Zaremba}.} \bibinfo{year}{2016}\natexlab{}.
\newblock \showarticletitle{Openai gym}.
\newblock \bibinfo{journal}{\emph{arXiv preprint arXiv:1606.01540}}
  (\bibinfo{year}{2016}).
\newblock


\bibitem[\protect\citeauthoryear{Choudhary, Gorla, and Orso}{Choudhary
  et~al\mbox{.}}{2015a}]%
        {androtest}
\bibfield{author}{\bibinfo{person}{Shauvik~Roy Choudhary},
  \bibinfo{person}{Alessandra Gorla}, {and} \bibinfo{person}{Alessandro Orso}.}
  \bibinfo{year}{2015}\natexlab{a}.
\newblock \showarticletitle{Automated test input generation for android: Are we
  there yet?(e)}. In \bibinfo{booktitle}{\emph{2015 30th IEEE/ACM International
  Conference on Automated Software Engineering (ASE)}}. IEEE,
  \bibinfo{pages}{429--440}.
\newblock


\bibitem[\protect\citeauthoryear{Choudhary, Gorla, and Orso}{Choudhary
  et~al\mbox{.}}{2015b}]%
        {automated}
\bibfield{author}{\bibinfo{person}{Shauvik~Roy Choudhary},
  \bibinfo{person}{Alessandra Gorla}, {and} \bibinfo{person}{Alessandro Orso}.}
  \bibinfo{year}{2015}\natexlab{b}.
\newblock \showarticletitle{Automated test input generation for android: Are we
  there yet?(e)}. In \bibinfo{booktitle}{\emph{2015 30th IEEE/ACM International
  Conference on Automated Software Engineering (ASE)}}. IEEE,
  \bibinfo{pages}{429--440}.
\newblock


\bibitem[\protect\citeauthoryear{Clark and Amodei}{Clark and Amodei}{2016}]%
        {faulty_rewards}
\bibfield{author}{\bibinfo{person}{Jack Clark} {and} \bibinfo{person}{Dario
  Amodei}.} \bibinfo{year}{2016}\natexlab{}.
\newblock \bibinfo{title}{Faulty Reward Functions in the Wild}.
\newblock
\newblock
\urldef\tempurl%
\url{https://openai.com/blog/faulty-reward-functions/}
\showURL{%
Retrieved January 4, 2021 from \tempurl}


\bibitem[\protect\citeauthoryear{Dong, Bohme, Cojocaru, and Roychoudhury}{Dong
  et~al\mbox{.}}{2020a}]%
        {timeMachinesys}
\bibfield{author}{\bibinfo{person}{Zhen Dong}, \bibinfo{person}{Marcel Bohme},
  \bibinfo{person}{Lucia Cojocaru}, {and} \bibinfo{person}{Abhik
  Roychoudhury}.} \bibinfo{year}{2020}\natexlab{a}.
\newblock \bibinfo{title}{Github Repository: TimeMachine}.
\newblock
\newblock
\urldef\tempurl%
\url{https://github.com/DroidTest/TimeMachine/blob/master/fuzzingandroid/sys_event_generator/sys_event.py}
\showURL{%
Retrieved November 20, 2020 from \tempurl}


\bibitem[\protect\citeauthoryear{Dong, Bohme, Cojocaru, and Roychoudhury}{Dong
  et~al\mbox{.}}{2020b}]%
        {dong2020timetravel}
\bibfield{author}{\bibinfo{person}{Zhen Dong}, \bibinfo{person}{Marcel Bohme},
  \bibinfo{person}{Lucia Cojocaru}, {and} \bibinfo{person}{Abhik
  Roychoudhury}.} \bibinfo{year}{2020}\natexlab{b}.
\newblock \showarticletitle{Time-travel Testing of Android Apps}. In
  \bibinfo{booktitle}{\emph{Proceedings of the 42nd International Conference on
  Software Engineering (ICSE)}}. \bibinfo{pages}{481--492}.
\newblock


\bibitem[\protect\citeauthoryear{Fujimoto, van Hoof, and Meger}{Fujimoto
  et~al\mbox{.}}{2018}]%
        {td3}
\bibfield{author}{\bibinfo{person}{Scott Fujimoto}, \bibinfo{person}{Herke van
  Hoof}, {and} \bibinfo{person}{David Meger}.} \bibinfo{year}{2018}\natexlab{}.
\newblock \showarticletitle{Addressing function approximation error in
  actor-critic methods}.
\newblock \bibinfo{journal}{\emph{arXiv preprint arXiv:1802.09477}}
  (\bibinfo{year}{2018}).
\newblock


\bibitem[\protect\citeauthoryear{Gao, Tan, Dong, and Roychoudhury}{Gao
  et~al\mbox{.}}{2018}]%
        {gao2018android}
\bibfield{author}{\bibinfo{person}{Xiang Gao}, \bibinfo{person}{Shin~Hwei Tan},
  \bibinfo{person}{Zhen Dong}, {and} \bibinfo{person}{Abhik Roychoudhury}.}
  \bibinfo{year}{2018}\natexlab{}.
\newblock \showarticletitle{Android testing via synthetic symbolic execution}.
  In \bibinfo{booktitle}{\emph{2018 33rd IEEE/ACM International Conference on
  Automated Software Engineering (ASE)}}. IEEE, \bibinfo{pages}{419--429}.
\newblock


\bibitem[\protect\citeauthoryear{Google}{Google}{2019}]%
        {broadcasts}
\bibfield{author}{\bibinfo{person}{Google}.} \bibinfo{year}{2019}\natexlab{}.
\newblock \bibinfo{title}{Broadcasts}.
\newblock
\newblock
\urldef\tempurl%
\url{https://developer.android.com/guide/components/broadcasts}
\showURL{%
Retrieved October 20, 2020 from \tempurl}


\bibitem[\protect\citeauthoryear{Google}{Google}{2020a}]%
        {emulator}
\bibfield{author}{\bibinfo{person}{Google}.} \bibinfo{year}{2020}\natexlab{a}.
\newblock \bibinfo{title}{Android Emulator}.
\newblock
\newblock
\urldef\tempurl%
\url{https://developer.android.com/studio/run/emulator/}
\showURL{%
Retrieved October 20, 2020 from \tempurl}


\bibitem[\protect\citeauthoryear{Google}{Google}{2020b}]%
        {api25}
\bibfield{author}{\bibinfo{person}{Google}.} \bibinfo{year}{2020}\natexlab{b}.
\newblock \bibinfo{title}{System-Level events API 25}.
\newblock
\newblock
\urldef\tempurl%
\url{https://cs.android.com/android/platform/superproject/+/android-7.1.2_r36:frameworks/base/core/res/AndroidManifest.xml}
\showURL{%
Retrieved December 28, 2020 from \tempurl}


\bibitem[\protect\citeauthoryear{Google}{Google}{2020c}]%
        {Monkey}
\bibfield{author}{\bibinfo{person}{Google}.} \bibinfo{year}{2020}\natexlab{c}.
\newblock \bibinfo{title}{UI/Application Exerciser Monkey}.
\newblock
\newblock
\urldef\tempurl%
\url{https://developer.android.com/studio/test/monkey}
\showURL{%
Retrieved November 10, 2020 from \tempurl}


\bibitem[\protect\citeauthoryear{Gu, Sun, Ma, Cao, Xu, Yao, Zhang, Lu, and
  Su}{Gu et~al\mbox{.}}{2019}]%
        {gu2019practical}
\bibfield{author}{\bibinfo{person}{Tianxiao Gu}, \bibinfo{person}{Chengnian
  Sun}, \bibinfo{person}{Xiaoxing Ma}, \bibinfo{person}{Chun Cao},
  \bibinfo{person}{Chang Xu}, \bibinfo{person}{Yuan Yao},
  \bibinfo{person}{Qirun Zhang}, \bibinfo{person}{Jian Lu}, {and}
  \bibinfo{person}{Zhendong Su}.} \bibinfo{year}{2019}\natexlab{}.
\newblock \showarticletitle{Practical GUI testing of Android applications via
  model abstraction and refinement}. In \bibinfo{booktitle}{\emph{2019 IEEE/ACM
  41st International Conference on Software Engineering (ICSE)}}. IEEE,
  \bibinfo{pages}{269--280}.
\newblock


\bibitem[\protect\citeauthoryear{Haarnoja, Zhou, Abbeel, and Levine}{Haarnoja
  et~al\mbox{.}}{2018}]%
        {sac}
\bibfield{author}{\bibinfo{person}{Tuomas Haarnoja}, \bibinfo{person}{Aurick
  Zhou}, \bibinfo{person}{Pieter Abbeel}, {and} \bibinfo{person}{Sergey
  Levine}.} \bibinfo{year}{2018}\natexlab{}.
\newblock \showarticletitle{Soft actor-critic: Off-policy maximum entropy deep
  reinforcement learning with a stochastic actor}.
\newblock \bibinfo{journal}{\emph{arXiv preprint arXiv:1801.01290}}
  (\bibinfo{year}{2018}).
\newblock


\bibitem[\protect\citeauthoryear{Hill, Raffin, Ernestus, Gleave, Kanervisto,
  Traore, Dhariwal, Hesse, Klimov, Nichol, Plappert, Radford, Schulman, Sidor,
  and Wu}{Hill et~al\mbox{.}}{2018}]%
        {stable-baselines}
\bibfield{author}{\bibinfo{person}{Ashley Hill}, \bibinfo{person}{Antonin
  Raffin}, \bibinfo{person}{Maximilian Ernestus}, \bibinfo{person}{Adam
  Gleave}, \bibinfo{person}{Anssi Kanervisto}, \bibinfo{person}{Rene Traore},
  \bibinfo{person}{Prafulla Dhariwal}, \bibinfo{person}{Christopher Hesse},
  \bibinfo{person}{Oleg Klimov}, \bibinfo{person}{Alex Nichol},
  \bibinfo{person}{Matthias Plappert}, \bibinfo{person}{Alec Radford},
  \bibinfo{person}{John Schulman}, \bibinfo{person}{Szymon Sidor}, {and}
  \bibinfo{person}{Yuhuai Wu}.} \bibinfo{year}{2018}\natexlab{}.
\newblock \bibinfo{title}{Stable Baselines}.
\newblock \bibinfo{howpublished}{https://github.com/hill-a/stable-baselines}.
\newblock


\bibitem[\protect\citeauthoryear{Holm}{Holm}{1979}]%
        {Holm}
\bibfield{author}{\bibinfo{person}{Sture Holm}.}
  \bibinfo{year}{1979}\natexlab{}.
\newblock \showarticletitle{A simple sequentially rejective multiple test
  procedure}.
\newblock \bibinfo{journal}{\emph{Scandinavian journal of statistics}}
  (\bibinfo{year}{1979}), \bibinfo{pages}{65--70}.
\newblock


\bibitem[\protect\citeauthoryear{Koroglu, Sen, Muslu, Mete, Ulker, Tanriverdi,
  and Donmez}{Koroglu et~al\mbox{.}}{2018}]%
        {koroglu2018qbe}
\bibfield{author}{\bibinfo{person}{Yavuz Koroglu}, \bibinfo{person}{Alper Sen},
  \bibinfo{person}{Ozlem Muslu}, \bibinfo{person}{Yunus Mete},
  \bibinfo{person}{Ceyda Ulker}, \bibinfo{person}{Tolga Tanriverdi}, {and}
  \bibinfo{person}{Yunus Donmez}.} \bibinfo{year}{2018}\natexlab{}.
\newblock \showarticletitle{QBE: QLearning-based exploration of android
  applications}. In \bibinfo{booktitle}{\emph{2018 IEEE 11th International
  Conference on Software Testing, Verification and Validation (ICST)}}. IEEE,
  \bibinfo{pages}{105--115}.
\newblock


\bibitem[\protect\citeauthoryear{Lee}{Lee}{2009}]%
        {ptolemy}
\bibfield{author}{\bibinfo{person}{Edward~A. Lee}.}
  \bibinfo{year}{2009}\natexlab{}.
\newblock \bibinfo{booktitle}{\emph{Finite State Machines and Modal Models in
  Ptolemy II}}.
\newblock \bibinfo{type}{{T}echnical {R}eport} UCB/EECS-2009-151.
  \bibinfo{institution}{EECS Department, University of California, Berkeley}.
\newblock
\urldef\tempurl%
\url{http://www2.eecs.berkeley.edu/Pubs/TechRpts/2009/EECS-2009-151.html}
\showURL{%
\tempurl}


\bibitem[\protect\citeauthoryear{Li}{Li}{2017}]%
        {li2017deep}
\bibfield{author}{\bibinfo{person}{Yuxi Li}.} \bibinfo{year}{2017}\natexlab{}.
\newblock \showarticletitle{Deep reinforcement learning: An overview}.
\newblock \bibinfo{journal}{\emph{arXiv preprint arXiv:1701.07274}}
  (\bibinfo{year}{2017}).
\newblock


\bibitem[\protect\citeauthoryear{Li, Yang, Guo, and Chen}{Li
  et~al\mbox{.}}{2019}]%
        {li2019humanoid}
\bibfield{author}{\bibinfo{person}{Yuanchun Li}, \bibinfo{person}{Ziyue Yang},
  \bibinfo{person}{Yao Guo}, {and} \bibinfo{person}{Xiangqun Chen}.}
  \bibinfo{year}{2019}\natexlab{}.
\newblock \showarticletitle{Humanoid: a deep learning-based approach to
  automated black-box Android app testing}. In \bibinfo{booktitle}{\emph{2019
  34th IEEE/ACM International Conference on Automated Software Engineering
  (ASE)}}. IEEE, \bibinfo{pages}{1070--1073}.
\newblock


\bibitem[\protect\citeauthoryear{Lillicrap, Hunt, Pritzel, Heess, Erez, Tassa,
  Silver, and Wierstra}{Lillicrap et~al\mbox{.}}{2015}]%
        {DDPG}
\bibfield{author}{\bibinfo{person}{Timothy~P Lillicrap},
  \bibinfo{person}{Jonathan~J Hunt}, \bibinfo{person}{Alexander Pritzel},
  \bibinfo{person}{Nicolas Heess}, \bibinfo{person}{Tom Erez},
  \bibinfo{person}{Yuval Tassa}, \bibinfo{person}{David Silver}, {and}
  \bibinfo{person}{Daan Wierstra}.} \bibinfo{year}{2015}\natexlab{}.
\newblock \showarticletitle{Continuous control with deep reinforcement
  learning}.
\newblock \bibinfo{journal}{\emph{arXiv preprint arXiv:1509.02971}}
  (\bibinfo{year}{2015}).
\newblock


\bibitem[\protect\citeauthoryear{Machiry, Tahiliani, and Naik}{Machiry
  et~al\mbox{.}}{2013}]%
        {machiry2013dynodroid}
\bibfield{author}{\bibinfo{person}{Aravind Machiry}, \bibinfo{person}{Rohan
  Tahiliani}, {and} \bibinfo{person}{Mayur Naik}.}
  \bibinfo{year}{2013}\natexlab{}.
\newblock \showarticletitle{Dynodroid: An input generation system for android
  apps}. In \bibinfo{booktitle}{\emph{Proceedings of the 2013 9th Joint Meeting
  on Foundations of Software Engineering}}. \bibinfo{pages}{224--234}.
\newblock


\bibitem[\protect\citeauthoryear{Mahmood, Mirzaei, and Malek}{Mahmood
  et~al\mbox{.}}{2014}]%
        {mahmood2014evodroid}
\bibfield{author}{\bibinfo{person}{Riyadh Mahmood}, \bibinfo{person}{Nariman
  Mirzaei}, {and} \bibinfo{person}{Sam Malek}.}
  \bibinfo{year}{2014}\natexlab{}.
\newblock \showarticletitle{Evodroid: Segmented evolutionary testing of android
  apps}. In \bibinfo{booktitle}{\emph{Proceedings of the 22nd ACM SIGSOFT
  International Symposium on Foundations of Software Engineering}}.
  \bibinfo{pages}{599--609}.
\newblock


\bibitem[\protect\citeauthoryear{Mao, Harman, and Jia}{Mao
  et~al\mbox{.}}{2016}]%
        {mao2016sapienz}
\bibfield{author}{\bibinfo{person}{Ke Mao}, \bibinfo{person}{Mark Harman},
  {and} \bibinfo{person}{Yue Jia}.} \bibinfo{year}{2016}\natexlab{}.
\newblock \showarticletitle{Sapienz: Multi-objective automated testing for
  Android applications}. In \bibinfo{booktitle}{\emph{Proceedings of the 25th
  International Symposium on Software Testing and Analysis}}.
  \bibinfo{pages}{94--105}.
\newblock


\bibitem[\protect\citeauthoryear{Marc R.~Hoffmann}{Marc R.~Hoffmann}{2020}]%
        {jacoco}
\bibfield{author}{\bibinfo{person}{Evgeny~Mandrikov Marc R.~Hoffmann,
  Brock~Janiczak}.} \bibinfo{year}{2020}\natexlab{}.
\newblock \bibinfo{title}{Jacoco Code Coverage}.
\newblock
\newblock
\urldef\tempurl%
\url{https://www.eclemma.org/jacoco/}
\showURL{%
Retrieved November 15, 2020 from \tempurl}


\bibitem[\protect\citeauthoryear{Mariani, Pezze, Riganelli, and
  Santoro}{Mariani et~al\mbox{.}}{2012}]%
        {mariani2012autoblacktest}
\bibfield{author}{\bibinfo{person}{Leonardo Mariani}, \bibinfo{person}{Mauro
  Pezze}, \bibinfo{person}{Oliviero Riganelli}, {and} \bibinfo{person}{Mauro
  Santoro}.} \bibinfo{year}{2012}\natexlab{}.
\newblock \showarticletitle{Autoblacktest: Automatic black-box testing of
  interactive applications}. In \bibinfo{booktitle}{\emph{2012 IEEE Fifth
  International Conference on Software Testing, Verification and Validation}}.
  IEEE, \bibinfo{pages}{81--90}.
\newblock


\bibitem[\protect\citeauthoryear{Mnih, Kavukcuoglu, Silver, Graves, Antonoglou,
  Wierstra, and Riedmiller}{Mnih et~al\mbox{.}}{2013}]%
        {DQN}
\bibfield{author}{\bibinfo{person}{Volodymyr Mnih}, \bibinfo{person}{Koray
  Kavukcuoglu}, \bibinfo{person}{David Silver}, \bibinfo{person}{Alex Graves},
  \bibinfo{person}{Ioannis Antonoglou}, \bibinfo{person}{Daan Wierstra}, {and}
  \bibinfo{person}{Martin Riedmiller}.} \bibinfo{year}{2013}\natexlab{}.
\newblock \showarticletitle{Playing atari with deep reinforcement learning}.
\newblock \bibinfo{journal}{\emph{arXiv preprint arXiv:1312.5602}}
  (\bibinfo{year}{2013}).
\newblock


\bibitem[\protect\citeauthoryear{Pan, Huang, Wang, Zhang, and Li}{Pan
  et~al\mbox{.}}{2020}]%
        {qtest}
\bibfield{author}{\bibinfo{person}{Minxue Pan}, \bibinfo{person}{An Huang},
  \bibinfo{person}{Guoxin Wang}, \bibinfo{person}{Tian Zhang}, {and}
  \bibinfo{person}{Xuandong Li}.} \bibinfo{year}{2020}\natexlab{}.
\newblock \showarticletitle{Reinforcement learning based curiosity-driven
  testing of Android applications}. In \bibinfo{booktitle}{\emph{Proceedings of
  the 29th ACM SIGSOFT International Symposium on Software Testing and
  Analysis}}. \bibinfo{pages}{153--164}.
\newblock


\bibitem[\protect\citeauthoryear{Riedmiller}{Riedmiller}{2005}]%
        {riedmiller2005neural}
\bibfield{author}{\bibinfo{person}{Martin Riedmiller}.}
  \bibinfo{year}{2005}\natexlab{}.
\newblock \showarticletitle{Neural fitted Q iteration--first experiences with a
  data efficient neural reinforcement learning method}. In
  \bibinfo{booktitle}{\emph{European Conference on Machine Learning}}.
  Springer, \bibinfo{pages}{317--328}.
\newblock


\bibitem[\protect\citeauthoryear{Silver}{Silver}{2014}]%
        {DPG}
\bibfield{author}{\bibinfo{person}{et~al. Silver, David}.}
  \bibinfo{year}{2014}\natexlab{}.
\newblock \showarticletitle{Deterministic Policy Gradient Algorithms}.
\newblock  (\bibinfo{year}{2014}).
\newblock


\bibitem[\protect\citeauthoryear{Su, Meng, Chen, Wu, Yang, Yao, Pu, Liu, and
  Su}{Su et~al\mbox{.}}{[n.d.]}]%
        {DBLP:conf/sigsoft/SuMCWYYPLS17}
\bibfield{author}{\bibinfo{person}{Ting Su}, \bibinfo{person}{Guozhu Meng},
  \bibinfo{person}{Yuting Chen}, \bibinfo{person}{Ke Wu},
  \bibinfo{person}{Weiming Yang}, \bibinfo{person}{Yao Yao},
  \bibinfo{person}{Geguang Pu}, \bibinfo{person}{Yang Liu}, {and}
  \bibinfo{person}{Zhendong Su}.} \bibinfo{year}{[n.d.]}\natexlab{}.
\newblock \showarticletitle{Guided, stochastic model-based {GUI} testing of
  Android apps}. In \bibinfo{booktitle}{\emph{Proceedings of the 2017 11th
  Joint Meeting on Foundations of Software Engineering, {ESEC/FSE} 2017,
  Paderborn, Germany, September 4-8, 2017}}. \bibinfo{pages}{245--256}.
\newblock


\bibitem[\protect\citeauthoryear{Su, Meng, Chen, Wu, Yang, Yao, Pu, Liu, and
  Su}{Su et~al\mbox{.}}{2017}]%
        {api19}
\bibfield{author}{\bibinfo{person}{Ting Su}, \bibinfo{person}{Guozhu Meng},
  \bibinfo{person}{Yuting Chen}, \bibinfo{person}{Ke Wu},
  \bibinfo{person}{Weiming Yang}, \bibinfo{person}{Yao Yao},
  \bibinfo{person}{Geguang Pu}, \bibinfo{person}{Yang Liu}, {and}
  \bibinfo{person}{Zhendong Su}.} \bibinfo{year}{2017}\natexlab{}.
\newblock \bibinfo{title}{System-Level events API 19}.
\newblock
\newblock
\urldef\tempurl%
\url{https://sites.google.com/site/stoat2017/evaluation/stoat-s-system-level-events}
\showURL{%
Retrieved December 28, 2020 from \tempurl}


\bibitem[\protect\citeauthoryear{Sutton}{Sutton}{2014}]%
        {sutton}
\bibfield{author}{\bibinfo{person}{Sutton}.} \bibinfo{year}{2014}\natexlab{}.
\newblock \bibinfo{booktitle}{\emph{Reinforcement Learning: An Introduction}}.
\newblock


\bibitem[\protect\citeauthoryear{Taylor and Martinovic}{Taylor and
  Martinovic}{2017}]%
        {root_apps}
\bibfield{author}{\bibinfo{person}{Vincent~F Taylor} {and}
  \bibinfo{person}{Ivan Martinovic}.} \bibinfo{year}{2017}\natexlab{}.
\newblock \showarticletitle{Short paper: A longitudinal study of financial apps
  in the Google Play store}. In \bibinfo{booktitle}{\emph{International
  Conference on Financial Cryptography and Data Security}}. Springer,
  \bibinfo{pages}{302--309}.
\newblock


\bibitem[\protect\citeauthoryear{Watkins and Dayan}{Watkins and Dayan}{1992}]%
        {watkins1992q}
\bibfield{author}{\bibinfo{person}{Christopher~JCH Watkins} {and}
  \bibinfo{person}{Peter Dayan}.} \bibinfo{year}{1992}\natexlab{}.
\newblock \showarticletitle{Q-learning}.
\newblock \bibinfo{journal}{\emph{Machine learning}} \bibinfo{volume}{8},
  \bibinfo{number}{3-4} (\bibinfo{year}{1992}), \bibinfo{pages}{279--292}.
\newblock


\end{thebibliography}
\newpage
\section{Appendix}

\subsection{Study 1- Deep RL}

Certain parameters of the algorithms are omitted for simplicity, and can be found in the documentation of Stable Baselines \cite{stable-baselines}.

\textbf{Control Policy:}

Deep RL algorithms rely on a policy based on a MLP composed of 2 layers and 64 neurons each.
\noindent
\textbf{Learning rates:} 
\begin{itemize}
    \item DDPG: 0.0001
    \item SAC: 0.0003
    \item TD3: 0.0003
\end{itemize}

\begin{table}[h]
\begin{tabularx}{0.67\textwidth}{l|llllllll}
\textbf{DDPG} &  &  &  &  &  &  &  &  \\
\textbf{Configuration} & \textbf{1} & \textbf{2} & \textbf{3} & \textbf{4} & \textbf{5} & \textbf{6} & \textbf{7} & \textbf{8} \\ \hline
random\_exploration & 0.5 & 0.5 & 0.6 & 0.6 & 0.7 & 0.7 & 0.8 & 0.8 \\
nb\_train\_steps & 5 & 25 & 5 & 25 & 5 & 25 & 5 & 25
\end{tabularx}
\end{table}

\begin{table}[h]
\begin{tabularx}{0.67\textwidth}{l|llllllll}
\textbf{TD3} &  &  &  &  &  &  &  &  \\
\textbf{Configuration} & \textbf{1} & \textbf{2} & \textbf{3} & \textbf{4} & \textbf{5} & \textbf{6} & \textbf{7} & \textbf{8} \\ \hline
random\_exploration & 0.5 & 0.5 & 0.6 & 0.6 & 0.7 & 0.7 & 0.8 & 0.8 \\
train\_frequency & 25 & 100 & 25 & 100 & 25 & 100 & 25 & 100
\end{tabularx}
\end{table}

\begin{table}[h]
\begin{tabularx}{0.67\textwidth}{l|llllllll}
\textbf{SAC} &  &  &  &  &  &  &  &  \\
\textbf{Configuration} & \textbf{1} & \textbf{2} & \textbf{3} & \textbf{4} & \textbf{5} & \textbf{6} & \textbf{7} & \textbf{8} \\ \hline
target\_update\_interval & 1 & 1 & 2 & 2 & 5 & 5 & 10 & 10 \\
train\_frequency & 1 & 5 & 1 & 5 & 1 & 5 & 1 & 5
\end{tabularx}
\end{table}

\subsection{Study 1- Tabular RL}

\begin{table}[h]
\begin{tabularx}{0.72\textwidth}{l|llllllll}
\textbf{Q-Learning} &  &  &  &  &  &  &  &  \\
\textbf{Configuration} & \textbf{1} & \textbf{2} & \textbf{3} & \textbf{4} & \textbf{5} & \textbf{6} & \textbf{7} & \textbf{8} \\ \hline
epsilon & 0.5 & 0.6 & 0.7 & 0.8 & 0.5 & 0.6 & 0.7 & 0.8 \\
gamma & 0.99 & 0.99 & 0.99 & 0.99 & 0.9 & 0.9 & 0.9 & 0.9 \\
alpha & 0.628 & 0.628 & 0.628 & 0.628 & 0.628 & 0.628 & 0.628 & 0.628\\
\end{tabularx}
\end{table}

\subsection{Study 2}

Some parameters used in Study 2 have been selected from the results of Study 1 (see Section 6). 

\textbf{DDPG:}
\begin{itemize}
    \item Control Policy: MLP, 2 layers, 64 neurons
    \item Learning rate: 0.0001 
    \item nb\_train\_steps: 10
    \item random\_exploration: 0.7
\end{itemize}

\textbf{SAC:}
\begin{itemize}
    \item Control Policy: MLP, 2 layers, 64 neurons
    \item Learning rate: 0.0003
    \item train\_freq: 5
    \item target\_update\_interval: 10
\end{itemize}

\textbf{TD3:}
\begin{itemize}
    \item Control Policy: MLP, 2 layers, 64 neurons
    \item Learning rate: 0.0003
    \item train\_freq: 10
    \item random\_exploration: 0.8
\end{itemize}

\textbf{Q-Learning:}
\begin{itemize}
    \item epsilon: 0.8
    \item gamma: 0.9
    \item alpha: 0.628
\end{itemize}

\end{document}